\documentclass[
reprint,
superscriptaddress,%groupedaddress,%unsortedaddress,%runinaddress,
nofootinbib, %nobibnotes, %bibnotes,
 amsmath,amssymb,
 aps
]{revtex4-1}

\usepackage{graphicx}
\usepackage{bm}
\usepackage{hyperref}
\usepackage{color}
\usepackage{graphicx}

\newcommand{\beq}{\begin{equation}}
\newcommand{\eeq}{\end{equation}}
\def\bea{\begin{eqnarray}}
\def\eea{\end{eqnarray}}
\def\beal{\begin{align}}
\def\eal{\end{align}}
\newcommand{\bei}{\begin{itemize}}
\newcommand{\eei}{\end{itemize}}
\newcommand{\bmat}{\begin{matrix}}
\newcommand{\emat}{\end{matrix}}
\DeclareMathOperator{\tr}{Tr}

\newcommand{\Fig}[1]{Fig.~\ref{#1}}
\newcommand{\Eq}[1]{Eq.~(\ref{#1})}
\newcommand{\Eqs}[2]{Eqs.~(\ref{#1}) and (\ref{#2})}
\newcommand{\Sec}[1]{Sec.~\ref{#1}}
\newcommand{\App}[1]{Appendix~\ref{#1}}

\def\={\,=\,}
\def\+{\,+\,}
\def\-{\,-\,}

\def\Msun{M_\odot}
\def\MNFW{M_\text{NFW}}
\def\MBBH{M_\text{BBH}}
\def\Mvir{M_\text{vir}}
\def\NFW{\text{NFW}}

\def\Scrit{\Sigma_\text{crit}}

\def\Thatd{\hat{T}_d}

\begin{document}

\title{Small-scale shear: Peeling off diffuse subhalos with gravitational waves}

\author{Han Gil Choi}
 \email{alivespace@snu.ac.kr}
 \affiliation{Center for Theoretical Physics, Department of Physics and Astronomy, Seoul National University, Seoul 08826, Korea}

\author{Chanung Park}
 \email{tcwpark070497@snu.ac.kr}
 \affiliation{Center for Theoretical Physics, Department of Physics and Astronomy, Seoul National University, Seoul 08826, Korea}
 
\author{Sunghoon Jung}
 \email{sunghoonj@snu.ac.kr}
 \affiliation{Center for Theoretical Physics, Department of Physics and Astronomy, Seoul National University, Seoul 08826, Korea}
\affiliation{Astronomy Research Center, Seoul National University, Seoul 08826, Korea}

\date{\today}

\begin{abstract}
Subhalos at subgalactic scales ($M\lesssim 10^7 \Msun$ or $k\gtrsim 10^3 \,{\rm Mpc}^{-1}$) are pristine test beds of dark matter (DM). However, they are too small, diffuse and dark to be visible, in any existing observations. In this paper, we develop a complete formalism for weak and strong diffractive lensing, which can be used to probe such subhalos with chirping gravitational waves (GWs). Also, we show that Navarro-Frenk-White(NFW) subhalos in this mass range can indeed be detected individually, albeit at a rate of ${\cal O}(10)$ or less per year at BBO and others limited by small merger rates and large required SNR $\gtrsim 1/\gamma(r_0) \sim 10^3$. It becomes possible as NFW scale radii $r_0$ are of the right size comparable to the GW Fresnel length $r_F$, and unlike all existing probes, their lensing is more sensitive to lighter subhalos. 
Remarkably, our formalism further reveals that the frequency dependence of weak lensing (which is actually the detectable effect) is due to shear $\gamma$ at $r_F$. Not only is it consistent with an approximate scaling invariance, but it also offers a new way to measure the mass profile at a successively smaller scale of chirping $r_F \propto f^{-1/2}$. Meanwhile, strong diffraction that produces a blurred Einstein ring has a universal frequency dependence, allowing only detections. These are further demonstrated through semianalytic discussions of power-law profiles. 
Our developments for a single lens can be generalized and will promote diffractive lensing to a more concrete and promising physics in probing DM and small-scale structures.
\end{abstract}

\maketitle

\tableofcontents

%%%%%%%%%
\section{Introduction} \label{sec:intro}

Cold dark matter (CDM) hypothesis has successfully explained large-scale structures of the Universe, providing firm evidences of dark matter (DM). However, DM was never detected directly, and its properties in smaller scales are not yet well established. For decades, there has been missing satellites problem~\cite{Klypin:1999uc,Moore:1999nt}, where the observed number of luminous satellite galaxies is smaller than the prediction, although CDM predicts numerous structures --- (sub)halos --- at the subgalactic scale. Recently, it was argued that the completeness correction of star formation and detection efficiencies may resolve the discrepancy~\cite{Kim:2017iwr,Nadler:2020prv}. Many new observations of satellite galaxies since then by DES, PANSTRRS1, and Gaia~\cite{Nadler:2020prv,Banik:2019cza} are indeed making a better agreement down to (star-forming limit) $M \gtrsim 10^7 - 10^8 \Msun$.

This can be progressed much further by searching for DM subhalos below $10^7 - 10^8 \Msun$. Above all, such light subhalos do not harbor star formation~\cite{Bullock:2000wn,Bromm:2013iya}, hence they are free of baryonic physics and keep the pristine nature of DM.  Their number abundance, mass profile, and spatial distribution can all be important information of underlying DM models~\cite{Buckley:2017ijx}; warm, fuzzy, and axion DM models, as well as primordial black holes, predict significant deviations here~\cite{Hsueh:2019ynk,Gilman:2019nap,Nadler:2020prv}. They can also test CDM and the missing satellites problem more in more depth~\cite{Bullock:2017xww,Buckley:2017ijx}. Lastly, they might be around us in large numbers, affecting local direct detection.

However, the searches are challenging. First, they are dark (no stars). Second, they are diffuse in mass profile (no cooling and contraction by baryons) so that their gravitational effects are also suppressed; often too diffuse to produce strong-lensing images or Einstein arcs. In addition, the Navarro-Frenk-White(NFW) profile~\cite{Navarro:1995iw} is known to fit simulations and galactic-scale observations,
 but its validity at small scales is also not established. Core-vs-cusp may be another relevant problem about the central mass profile~\cite{Moore:1994yx,Flores:1994gz,Bullock:2017xww}.

Existing searches mainly rely on millilensing perturbations by subhalos.
When one of the strong-lensed images (of compact sources such as quasars) or an arc (of spatially extended sources such as galaxies) is near a subhalo, its flux, shape, location, and arrival time can be millilensing perturbed so that different from those of the other images or the other part of the arc~\cite{Mao:1997ek} (see also \cite{Chen:2003uu,Kochanek:2003zc,Metcalf:2004eh,Koopmans:2005nr,Sugai:2007ic,Xu:2011ru,Zackrisson:2009rc} and references therein). With excellent imaging and spatial resolution, this method can detect subhalos individually~\cite{Hezaveh:2012ai}, but only the heaviest ones down to $M \gtrsim 10^7 - 10^8\Msun$ for NFW~\cite{Nierenberg:2014cga} (and similarly for pseudo-Jaffe~\cite{Hezaveh:2016ltk,Asadi:2017,Nierenberg:2014cga,Fadely:2012}). The sensitivity is lower limited 
%partially by imaging spatial resolution and source sizes, but also 
inherently by profile diffuseness; NFW is so diffuse that millilensing cross section $\sigma_l \propto \MNFW^{2.5 - 5}$ scales rapidly with the mass, as estimated in \App{app:MNFWmilli}. (For comparison, compact DM can be probed down to very small masses with lensing~\cite{Niikura:2017zjd,Zumalacarregui:2017qqd,Munoz:2016tmg,Nakamura:1997sw,Jung:2017flg,Gould:1992,Katz:2018zrn,Nemiroff:1995ak,Jung:2019fcs,Dror:2019twh,Mondino:2020rkn}.)
Alternatively, a mass function~\cite{Metcalf:2001ap,Chiba:2001wk,Metcalf:2001es,Dalal:2001fq,Maccio:2005bj} or power spectrum~\cite{Hezaveh:2014aoa,Rivero:2017mao,Cyr-Racine:2018htu} can be extended below this range, through the collective or statistical effects of subhalos; the mass abundance inferred in this way also agrees better with CDM in the range $10^6 - 10^9\Msun$~\cite{Vegetti:2012mc,Vegetti:2018dly,Hsueh:2019ynk,Gilman:2019nap}. Thus, to search for individual (sub)halos below $10^7 \Msun$,\footnote{We note that searches using star kinematics~\cite{Erkal:2015kqa,Feldmann:2013hqa,Buschmann:2017ams,Bonaca:2018fek,Banik:2019cza} are also limited by $\gtrsim 10^8 \Msun$, similarly to the millilensing. Perhaps, it is partly because both rely on presumably similar size $\sim 10\%$ of gravitational perturbations. But the similar threshold of star-forming galaxies $\gtrsim 10^7 - 10^8\Msun$ might be a coincidence.} we need a very different method.

Recently, it has been proposed that diffractive lensing of chirping GWs can be used to probe relatively light pseudo-Jaffe subhalos of $M_{\rm vir} \lesssim 10^6 \Msun$ (more precisely, $M_E = 10^2 - 10^3\Msun$, where $M_E$ is the mass within the Einstein radius)~\cite{Dai:2018enj}. As will be discussed throughout this paper, the chirping GW is an ideal object to probe such subhalos; first because its Fresnel length coincides with the scale radii of such profiles~\cite{Takahashi:2005ug,Oguri:2020ldf}, the frequency chirping is so well under theoretical control that it can be used for precision measurements, and it is highly coherent, generated from an almost point source, retaining its diffraction pattern. 
The same physics has also been used to search for compact DM such as primordial black holes~\cite{Nakamura:1997sw,Jung:2017flg,Takahashi:2003ix,Lai:2018rto,Christian:2018vsi} and cosmic strings~\cite{Suyama:2005ez,Jung:2018kde}. 
These works have pioneered diffractive lensing near the Einstein radius, $r_E$, but NFW is more diffuse with essentially zero $r_E$ (see \Sec{sec:critical}). Not only is it difficult to calculate their diffractive lensing even numerically, but it is also not clear which scales are relevant and how strong the lensing will be.

\begin{figure}
\includegraphics[width=0.85\linewidth]{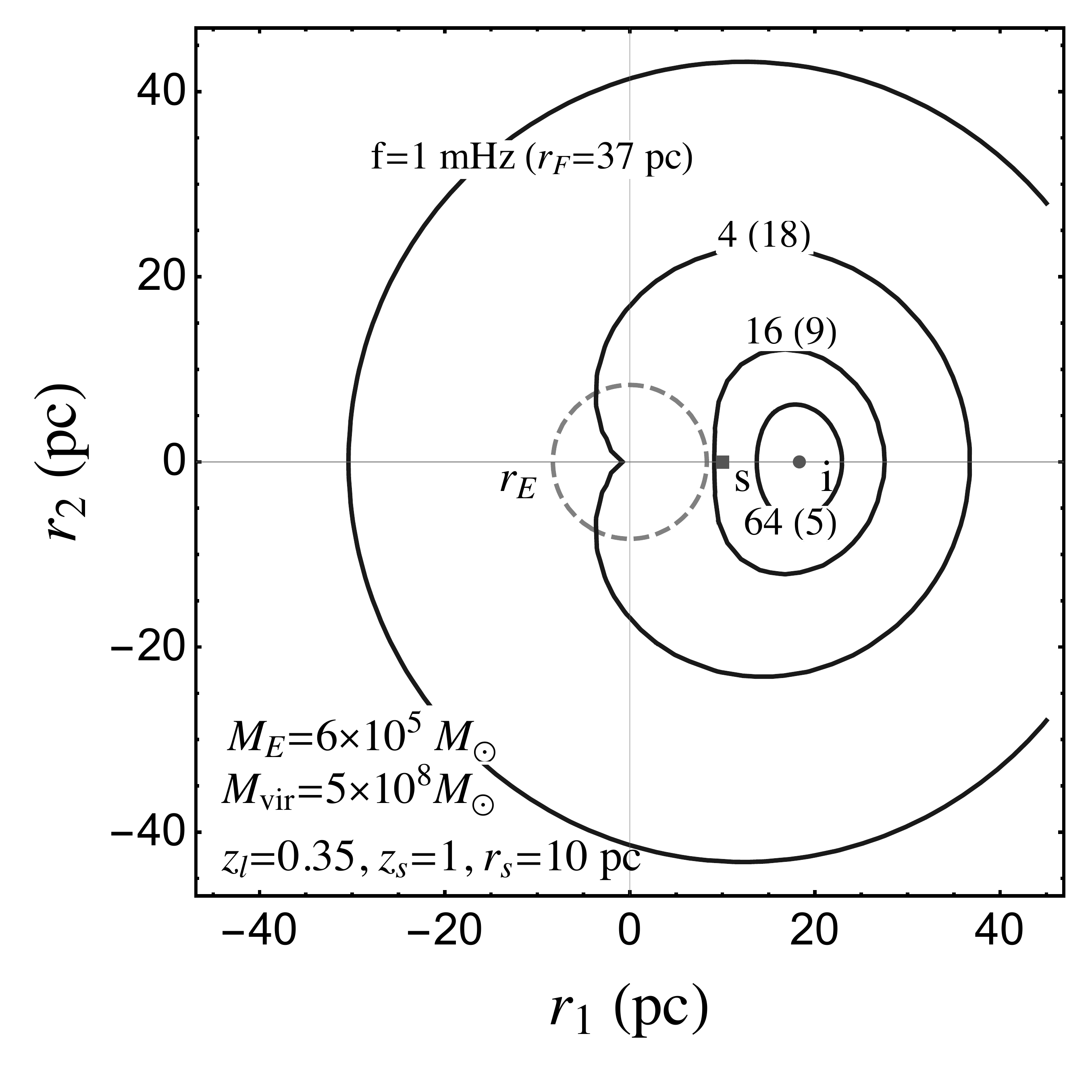}
\caption{ \label{fig:overview} 
Illustrating how the chirping GW detects a diffuse subhalo and successively peels off its profile. The solid circles with radii $\sim$ Fresenel length $r_F \propto 1/\sqrt{f}$ are the points on the lens plane being probed by the wave with frequency $f$ and are also where the phase difference with an image ``i'' is 1.
As the frequency chirps, the circle shrinks and the wave feels the mass distribution at successively smaller scales, hence frequency-dependent diffractive lensing is essentially due to shear. When $r_F \lesssim r_s$, the influence of the source ``s'' to the phase of the wave becomes non-negligible and the image ``i'' begins to be well located by the Fermat principle, hence geometric optics. Singular Isothermal Sphere (SIS) is used for illustration, where mass is densely distributed within the Einstein radius $r_E$ which is also a boundary between weak and strong diffraction. NFW is more diffuse with essentially zero $r_E$.
}
\end{figure}

In this paper, we develop a general formalism for diffractive lensing and work out the lensing of GW induced by a single NFW halo, both analytically and numerically (see Refs.~\cite{2019RMxAA,Guo:2020eqw} for some numerical results). GW diffraction has been already proposed to measure the matter power spectrum that includes NFW halos at small scales $1-10^4\Msun$~\cite{Takahashi:2005ug,Oguri:2020ldf} or solar-mass microlens populations~\cite{Diego:2019lcd,Mishra:2021xzz}. But focusing on an individual lens, we aim to assess the prospects of individual detection and profile measurements. Along this line, our formalism provides an easier description in terms of 2D potentials as well as a basic understanding of qualitatively different regimes of diffractive lensing. Some of the main underlying physics is illustrated in \Fig{fig:overview} and will be discussed throughout.

We start by developing general formalism in \Sec{sec:formalism}, then we solve NFW diffractive lensing in \Sec{sec:NFW}, introduce and quickly estimate the GW lensing detection in \Sec{sec:detection}, and present numerical results of detection prospects in \Sec{sec:results}. We demonstrate the application of our formalism to more general profiles in \Sec{sec:discussion}. We close by summarizing the results in \Sec{sec:summary}.

%%%%%%%%%
\section{Diffractive lensing formalism} \label{sec:formalism}

We develop general formalism for diffractive lensing.

%%%%%
\subsection{Lensing integral}

Gravitational lensing effects are captured generally by a complex amplification factor $F(f)$ as
\beq
    \tilde{h}_L(f) \= F(f) \tilde{h}(f),
\eeq
where $\tilde{h}$ ($\tilde{h}_L$) is an unlensed (lensed) waveform in the frequency $f$ domain. The amplification is calculated using the Kirchhoff path integral on the lens plane as~\cite{Nakamura:1999}
\bea \label{eq:physicalF} 
    F(f) \= \frac{f(1+z_l)}{ i d_\text{eff} }\int d^2\bm{r}\exp \left[i 2\pi f(1+z_l) T_d(\bm{r},\bm{r}_s)\right],
\eea
where $\bm{r}$ is the physical displacement on the lens plane with its origin at the center of the lens, $\bm{r}_s$ is the source position projected onto the lens plane, $T_d$ is the arrival-time difference between the deflected path passing $\bm{r}$ under the lens influence and a straight path in the absence of the lens, and $d_\text{eff}=d_l d_{ls}/d_s$ is the effective angular-diameter distance to the lens.

It is convenient to normalize dimensionful parameters by a characteristic length scale $r_0$
\bea \label{eq:dimlessF}
     F(w) \= \frac{w}{ 2\pi i }\int d^2\bm{x}\exp \left[ i w \hat{T}_d(\bm{x},\bm{x}_s)\right] \, ,
 \eea
where $\bm{x}=\bm{r}/r_0$, $\bm{x}_s=\bm{r}_s/r_0$, $\hat{T}_d=d_\text{eff}T_d/r_0^2$, and 
\bea \label{eq:w}
    w \, \equiv \, 2\pi f(1+z_l) \frac{r_0^2}{d_\text{eff}}
\eea
is the dimensionless frequency. The dimensionless time delay $\hat{T}_d$, also called the Fermat potential, is
\begin{equation}
    \hat{T}_d (\bm{x},\bm{x}_s) \= \frac{1}{2}|\bm{x}-\bm{x}_s|^2 - \psi(\bm{x}).
\label{eq:hatTd} \end{equation}
The first term denotes the geometric time delay and the second the Shapiro delay with dimensionless potential $\psi$ being the solution of two-dimensional Poisson equation
\begin{equation}
    \nabla^2_{\bm{x}} \psi \= 2 \kappa(\bm{x}) \=\frac{2\Sigma(\bm{x})}{\Scrit},
\label{eq:laplacian} \end{equation}
with the surface density projected onto the lens plane
\beq
\Sigma(\bm{x}) \= \int_{-\infty}^{\infty} dz \, \rho(z,\bm{x}), \quad \Scrit \= \frac{1}{ 4 \pi d_{\rm eff}}.
\eeq
The convergence $\kappa(\bm{x})$ is the normalized surface density, characterizing lensing strength.

The formalism so far is general and scale invariant. The normalization $r_0$ can be chosen to be any convenient scale of the lens. For example, the Einstein radius $r_E=\sqrt{4 M_E d_{\rm eff}}$ is a convenient choice of $r_0$ for a point-mass lens because its  $M_E$ is equal to the total mass $M$ so that $w = 8\pi M f$ is  a simple function of $M$. Thus, such a choice is often used for strong lensings (see \Sec{sec:discussion} for the usage for power-law profiles).

For diffuse lenses such as NFW, which rarely induce strong lensing, it is more intuitive and useful to rewrite $w$ in \Eq{eq:w} in terms of a new length scale $r_F$ such that
\beq \label{eq:w2}
    w \= 2\left(\frac{r_0}{r_F}\right)^2.
\eeq
The new scale defined as~\cite{Macquart:2004sh,Takahashi:2005ug}
\beq \label{eq:rF}
     r_F \, \equiv \, \sqrt{\frac{d_\text{eff}}{\pi f(1+z_l)}} 
    \, \simeq \, 1.76\, \text{pc} \, \sqrt{\frac{1}{1+z_l}\left(\frac{d_\text{eff}}{\text{Gpc}}\right)\left(\frac{\text{Hz}}{f}\right)},
\eeq
is equivalent to the Fresnel length of diffraction applied to lensing. We will use $r_F$ throughout this paper, discussing its meaning and usefulness in later sections.

Usual geometric-optics lensing is obtained for $w \hat{T}_d \gg 1$ from the stationary points of $\hat{T}_d$, hence the Fermat principle.

%%%%%
\subsection{Diffraction condition} \label{sec:diffcond}

Diffractive lensing (also called wave-optics lensing) is the lensing in the regime where the Fermat principle does not lead to clear discrete paths of waves from the Kirchhoff path integral. It is where the wave properties of a probe wave becomes relevant. This typically produces a single blurred image of a source when $r_s \gg r_E$. 
But for $r_s \lesssim r_E$, would-be multiple images may not be well resolved and interfere; such is also referred to as the wave-optics effect~\cite{Jung:2017flg,Katz:2018zrn,Jow:2020rcy,Takahashi:2003ix}.
In this subsection, we derive the conditions for diffractive lensing.

The Fermat principle applies when the phase oscillation among the paths passing different parts of the lens plane is rapid enough, i.e. $2\pi f  T_d \gg 1$ near $r_s$ in \Eq{eq:physicalF}. Thus, diffractive lensing occurs when, in terms of  $w$ in \Eq{eq:dimlessF}, 
\beq
w \hat{T}_d \, \simeq\, w \frac{x_s^2}{2} \,\lesssim\, 1,  
\label{eq:waveoptics} \eeq
where the approximate equality holds if $r_s \gg r_E$ so that the $\psi$ contribution to $\hat{T}_d$ in \Eq{eq:hatTd} is negligible compared to the geometric contribution. 

Diffractive lensing can also be understood by the analogy with single-slit experiment. The shadow of a slit is blurred when light rays propagating from opposite edges of the slit interfere weakly. This happens when the phase difference between them, $2\pi(\sqrt{a^2+d^2}-d)/\lambda\sim \pi a^2/(\lambda d)=(a/r_F)^2$, is small~\cite{Thorne:2017}; here, $a$, $d$, and $\lambda$ are the slit size, the distance between the slit and the screen, and the wavelength of incident light, respectively. In gravitational lensing, $a$ and $d$ are replaced by $ r_s$ (single-imaged cases) and $d_\text{eff}$, respectively. Thus, diffractive lensing occurs if
\beq
r_F^2 \,\gtrsim\, r_s^2,
\label{eq:waveoptics2} \eeq
which is equivalent to \Eq{eq:waveoptics} with the definition of $w$ in \Eq{eq:w2}. The condition in this form means that as chirping $r_F$ falls below $r_s$, the source becomes well located and only the lens mass profile near the source direction begins to matter; see \Fig{fig:overview} and \Sec{sec:strong}. $r_F$ is essentially an effective source size~\cite{Oguri:2020ldf}, within which effects are smeared/interfered out.

The diffraction picture is refined when $r_s \lesssim r_E$ (or, $r_s$ near any caustic) so that a lens system can have multiple images. The deflection potential $\psi$ now significantly contributes to $\Delta \hat{T}_d$. A more appropriate diffraction condition is $2\pi f \Delta T_{ij} \lesssim 1$ or $w \Delta \hat{T}_{ij} \lesssim 1$ (rather than \Eq{eq:waveoptics}),  
where $\Delta T_{ij}$ is the arrival-time difference between the $i$ th and $j$ th images~\cite{Takahashi:2003ix}. 
Since typically $\Delta T_{ij} \sim 4M_E = r_E^2/d_{\rm eff}$ (equivalently, $\Delta \hat{T}_{ij} \sim 1$ with $r_0=r_E$), the condition becomes $r_F^2 \gtrsim r_E^2$ [cf. \Eq{eq:waveoptics2}]. 
Applied to the point-mass lensing, the condition leads to a well-known interference relation $\lambda \gtrsim 2\pi R_{Sch}$ between the probe wavelength $\lambda$ and the lens Schwarzschild radius $R_{Sch} = 2M$, as $r_E^2/r_F^2 = (4 M d_\text{eff})/( \lambda d_\text{eff}/\pi) = 4\pi M/\lambda \lesssim 1$. Thus, this relation is nothing but the requirement for the wave to see the lens (or the slit in the single-slit analogy), or equivalently for the interference between multiple images to be relevant~\cite{Jung:2017flg,Katz:2018zrn,Jow:2020rcy,Takahashi:2003ix}.

Wave properties (hence, frequency dependencies) remain important inside $r_E$ up until $r_F \gtrsim 2\sqrt{r_E r_s}$. Consider $x_s \to 0$ near a caustic. The would-be multiple images have very small relative time delays, $\Delta \hat{T}_d = 2 x_E x_s + {\cal O}(x_s^2)$ (derived in \App{app:caustic}), as they are formed almost symmetrically around the corresponding critical lines (in this case, the Einstein radius $x_E$). Only if the frequency is very large, the resulting interference becomes so rapid that geometric optics is reached. Thus, diffraction continues well inside the Einstein radius until
\beq
w \, \lesssim \, \frac{1}{2 x_E x_s} \qquad \leftrightarrow \qquad r_F \,\gtrsim \, 2 \sqrt{r_E r_s}.
\label{eq:generalcond}
\eeq
Diffraction inside $r_E$ is strong lensing, and it produces a blurred Einstein ring, which becomes sharper as $r_F$ decreases toward this limit, eventually separating into clear images. 

In all, \Eq{eq:waveoptics} or (\ref{eq:waveoptics2}) is a relevant diffraction condition for NFW (\Sec{sec:critical}). But $r_E$ and strong diffractive lensing with \Eq{eq:generalcond} can also be relevant to general diffuse profiles (\Sec{sec:discussion}). In the next subsections, we formulate diffractive lensing and see how these physics arise.

%%%%%
\subsection{Formalism for weak diffractive lensing} \label{sec:diffraction}

We solve \Eq{eq:dimlessF} for weak diffractive lensing, in terms of much simpler 2D projected potentials. This formalism is applicable to any single lens profiles without symmetries. Weak lensing will be relevant to NFW.

In the diffraction regime $r_F \gtrsim r_s$, it is convenient to ignore $x_s$ (effectively, not well resolved) so that \Eq{eq:dimlessF} is rewritten as
\begin{align} \label{eq:Fshift}
    F(w) \,\simeq\, \frac{w}{2\pi i } \int d^2 \bm{x} \exp\left[i w \left(\frac{1}{2}|\bm{x}|^2 - \psi (\bm{x})-T_0\right)\right]\, .
\end{align}
$T_0$ is the overall time delay in the geometric-optics limit relative to the unlensed case; $F(w)$ now contains only the relative time delays among diffracted rays. We will see later what $T_0$ means for both the single- and multi-imaged cases.

For weak diffraction with small $\psi$ (more precisely, when the Shapiro delay is subdominant or $r_s \gtrsim r_E$), the Born approximation leads to the expansion
\begin{align} \label{eq:Fborn}
    F(w) \,\simeq \, 1- \frac{w^2}{2\pi } \int d^2 \bm{x} \, e^{ \frac{1}{2}i w|\bm{x}|^2} \left( \psi (\bm{x})-\psi(0)\right)\, ,
\end{align}
where $T_0 \simeq -\psi(0)$ for weak lensing. Using the integration by parts [with $i w x \, e^{i w x^2/2} = \frac{d}{dx}(e^{i w x^2/2})$], \Eq{eq:Fborn} can be written as 
\begin{align} \label{eq:Fconv}
    F(w) \, \simeq \, 1+ \frac{w}{i} \int_0^{\infty} dx \,x e^{ i w\frac{x^2}{2} } \, \overline{\kappa}(x)\, ,
\end{align}
where $\overline{\kappa}(x)$ is the mean convergence within the aperture of radius $x$ centered at $\bm{x}_s$ as \cite{Schneider:2006}
\begin{align} \label{eq:meanconv}
    \overline{\kappa}(x)& \,\equiv\, \frac{1}{\pi x^2} \int_{|\bm{x}'|<x}d^2\bm{x}' \kappa(\bm{x}')\notag\\
    &\= \frac{1}{2\pi x }\int_0^{2\pi} d\phi' \, \frac{\partial }{\partial x}\psi (x,\phi')\, 
\end{align}
with the lens-plane polar coordinate $(x, \phi)$.

Furthermore, important physics is contained in the frequency dependence of $F(w)$. By differentiating \Eq{eq:meanconv}, 
\begin{align}\label{eq:meantshear}
    \langle\gamma_t(x)\rangle \,\equiv\, \frac{1}{2\pi}\int_0^{2\pi}d\phi \, \gamma_t (x, \phi) \= -\frac{1}{2}\frac{d \overline{\kappa}(x)}{d \ln x}\,  ,
\end{align}
where $\gamma_t$ is the tangential shear
\begin{align} \label{eq:tshear}
    \gamma_t(x,\phi) \= \frac{1}{2}\left[\frac{1}{x}\frac{\partial \psi}{\partial x}-\frac{\partial^2 \psi}{\partial x^2}+\frac{1}{x^2}\frac{\partial^2 \psi }{\partial \phi^2}\right]\, .
\end{align}
Using \Eq{eq:meantshear}, the differentiation of \Eq{eq:Fconv} with respect to $\ln w$ can be written in terms of shear
\begin{align}\label{eq:dFshear}
    \frac{d F(w)}{d \ln w} \= \frac{w}{i} \int_0^{\infty} dx \,x e^{iw\frac{x^2}{2}}\langle\gamma_t(x)\rangle\, .
\end{align}

Finally and remarkably, although \Eqs{eq:Fconv}{eq:dFshear} are already new and insightful results of this work, they can be more usefully simplified as 
\begin{align}
    F(w) &\,\simeq \, 1+ \overline{\kappa}\left(\frac{1}{\sqrt{w}}e^{i\frac{\pi}{4}}\right) \label{eq:Fconv2}\\
    \frac{d F(w)}{d \ln w} &\,\simeq\, \left\langle \gamma_t\left(\frac{1}{\sqrt{w}}e^{i\frac{\pi}{4}}\right)\right\rangle\, , \label{eq:dFshear2}
\end{align}
in that the dominant support of the integral $\int_0^\infty dx \, x e^{i w x^2/2}$ is near $x=\frac{1}{\sqrt{w}} e^{i \pi/4}$, which can be obtained by rotating the half real-line integration by $e^{i\pi/4}$. The phase factor in the support is crucial to make this single region a dominant contributor. These are good approximations as long as $\overline{\kappa}(x)$ and $\langle \gamma_t(x) \rangle$ do not vary rapidly near the support. 

\Eqs{eq:Fconv2}{eq:dFshear2} are one of the new and main results of this paper. The fact that the complicated lensing integral is evaluated by much simpler 2D potentials is not only very convenient in estimating and understanding diffractive lensing, but also has various implications. Such utilities and implications will be  discussed and demonstrated throughout this paper.

\medskip
Before moving on, we discuss the formalism in more detail. First, $F(w)$ is a complex quantity, containing information on both amplification $|F(w)|$ and phase $\varphi(w)$ (or interferences). For small $\psi$, one can decompose as~\cite{Takahashi:2005ug}
\begin{align}
    |F(w)|  &\simeq  \text{Re}[F(w)] = 1+w\int_0^{\infty}dx \,x \sin \frac{wx^2}{2} \overline{\kappa}(x), \label{eq:Fabs}\\
    \varphi(w) &\simeq \text{Im}[F(w)] = -w\int_0^{\infty}dx \, x \cos \frac{wx^2}{2} \overline{\kappa}(x), \label{eq:Fphase}
\end{align}
and
\begin{align}
    \frac{d|F(w)|}{d\ln w} &\simeq \text{Re}\left[\frac{d F(w)}{d\ln w}\right]= w\int_0^{\infty}dx \,x \sin \frac{wx^2}{2} \langle\gamma_t(x)\rangle, \label{eq:dFabs}\\
    \frac{d\varphi(w)}{d\ln w} &\simeq \text{Im}\left[\frac{d F(w)}{d\ln w}\right]=- w\int_0^{\infty}dx \,x \cos \frac{wx^2}{2} \langle\gamma_t(x)\rangle\, \label{eq:dFphase}. 
\end{align}
The frequency dependencies of amplification and phase are of the same order and governed commonly by shear. Both physics must be utilized for detection and precision measurements.

%\medskip
Up to this point, no assumptions on $\psi$ were made except for its smallness. For the axisymmetric profiles considered in this paper,   the angular dependence is trivial so that 2D identities are simplified as
\begin{align}
    \overline{\kappa}(x) &\= \frac{1}{x}\psi'(x), \label{eq:aximeanconv}\\
    \langle\gamma_t(x)\rangle &\= \gamma(x) \= \frac{1}{2}\left[\frac{1}{x}\psi'(x)-\psi''(x)\right]\, . \label{eq:aximeanshear}
\end{align}
From here on, we will drop the subscript `t' for shear. Thus, we arrive at final formula for an axisymmetric lens
\begin{align}\label{eq:meankFapprx}
    F(w) &\simeq \, 1 +\frac{w}{i} \int_0^{\infty}dx \,x e^{i w \frac{x^2}{2}}\, \overline{\kappa}(x) \,\simeq\,  1 + \overline{\kappa}\left(\frac{1}{\sqrt{w}}e^{i\frac{\pi}{4}}\right), \\
    \frac{dF(w)}{d\ln w} &\simeq \, \frac{w}{i}\int_0^{\infty} d x \, x e^{i w \frac{x^2}{2}}\, \gamma(x) \,\simeq\,  \gamma \left(\frac{1}{\sqrt{w}}e^{i\frac{\pi}{4}}\right).   \label{eq:gammadFapprx}
\end{align}

%%%
\subsection{Shear as the origin of frequency dependence} \label{sec:shear}

The most remarkable meaning of \Eq{eq:dFshear2} or (\ref{eq:gammadFapprx}) is that the origin of the frequency dependence is (1) `shear' of a lens, and (2) at frequency-dependent $x \simeq 1/\sqrt{w}$ or $r \simeq r_F/\sqrt{2}$. 

Why does this make sense? 
Shear, defined in \Eq{eq:aximeanshear}, is produced from asymmetric mass distributions, hence distorting the shapes of background galaxies; it also reflects how steeply a profile varies at a given point. Consider the expression in the form
\beq
\gamma(x) \= \overline{\kappa}(x) - \kappa(x),
\label{eq:gammanew} \eeq 
derived from \Eqs{eq:aximeanconv}{eq:aximeanshear} and $ \kappa(x) = \frac{1}{2}\nabla^2 \psi(x) = \frac{1}{2} \left( \psi'(x)/x + \psi''(x) \right)$ for axisymmetric cases.
Note that $\overline{\kappa}(x)$ [hence $\gamma(x)$] does not necessarily vanish in the vicinity of a lens even though the density $\kappa(x) \propto \Sigma(x)$ may vanish there. So this form makes it clear that the variation of the potential is the one that produces shear, except at the spherically symmetric point (as a component of the Weyl conformal curvature tensor~\cite{Holz:1997ic,Wald:GRbook}). 

Further, \Eqs{eq:meankFapprx}{eq:gammadFapprx} are consistent with Gauss' theorem; gravitational effects must depend only on the enclosed mass. The enclosure boundary in our problem is given by the diffraction length scale $r_F \propto f^{-1/2}$. Thus, as the frequency grows, the boundary shrinks and the enclosed mass changes (see \Fig{fig:overview} for illustration). The change of lensing effects is a function of frequency and thus must be related to the variation of the mass or potential at the boundary, which is given by shear.

Nevertheless, geometric optics is frequency independent.
As $r_F \lesssim r_s$ or $2\sqrt{r_E r_s}$, the source is well resolved and the Fermat principle determines image properties solely from $\hat{T}_d$ in the narrow region around the image. (This will be further discussed in the next subsection.)
In reality, a mass profile may contain several substructures at various scales of their own small curvatures. If we probe this profile with a broad range of $r_F$, every time $r_F$ crosses this scale of a substructure, a wave-optics effect perturbing and correcting the image properties accounting for the substructure influence appears.

\Eq{eq:gammadFapprx} offers a new concrete way to measure the mass profile. The measurement of $dF(w)/d\ln w$ for a range of $w$ (even from a single GW event) can be directly translated to the measurement of the shear field $\gamma(x)$ for the corresponding length range; recall that $F(w)$ cannot be measured directly. Just as the shear field measured from galaxy shape distortions are used to measure the mass of a lens galaxy cluster, the shear field from GW diffraction (this time even with a single event) can tell the lens mass profile. 
In \Sec{sec:peelprofiles}, we apply our formalism to briefly  demonstrate this physics potential.

Practically, \Eq{eq:gammadFapprx} allows us to estimate diffractive lensing much more easily. The Kirchhoff integral is usually very difficult to calculate even numerically, but 2D projected potentials are much easier. In the following sections, we work out NFW diffractive lensing both analytically and numerically, not only confirming our formalism but also showing how readily one can estimate diffractive lensing.

%%%%%
\subsection{Complete formalism with strong diffraction} \label{sec:strong}

When the Einstein radius of a lens can be comparable to the $r_F$ of chirping GWs, strong diffractive lensing (which is qualitatively different from the weak diffractive lensing) must be taken into account. As derived in \Eq{eq:generalcond}, strong diffractive lensing occurs if $2\sqrt{r_E r_s} \lesssim r_F \lesssim r_E$. Given this condition, one can show that the main contributions to the lensing integral \Eq{eq:dimlessF} arise at $x \simeq x_E$, i.e., the Einstein ring. Using the stationary phase approximation at $x=x_E$, \Eq{eq:dimlessF} is evaluated as
\begin{align}\label{eq:strongdiff}
    F(f) \,\simeq\, i^{-\frac{1}{2}}x_E\sqrt{\frac{2\pi w }{1-\kappa(x_E)+\gamma(x_E)}}\, ,
\end{align}
which is again expressed in terms of $\kappa$ and $\gamma$; this time at $x=x_E$.

Interestingly, the frequency dependence $F(f)\propto w^{1/2}$ of strong diffractive lensing is universal to all axisymmetric lenses. This can be intuitively understood from the shape of an Einstein ring, which is produced since $r_s$ is negligible.
By the diffraction effect, the ring is blurred so that it looks like an annulus with thickness $\sim r_F$ and radius $\sim r_E$. Then, one can expect \Eq{eq:dimlessF} to be $F(f)\propto r_F^{-2} \times \text{(area of the annulus)}$ $\propto r_E r_F^{-1} = x_E \sqrt{w} $, and this is exactly as in \Eq{eq:strongdiff}. 

The situation is different in the weak diffraction regime, where $F(f)$ directly connects to the lens profile through $\overline{\kappa}$ and $\gamma$ at $x\simeq r_F$. What is the origin of the difference between the two diffraction regimes? It is due to the approximate scale invariance in the weak diffraction regime; no length scales up to weak gravitational potential $\psi$. In contrast to strong diffractive lensing, the weak lensing integral is dominated by a disk with radius $r_F$ centered at the origin. By a similar argument, one might expect $F(f) \propto r_F^{-2} \times r_F^2\propto \text{const}$, which looks at first inconsistent with \Eqs{eq:meankFapprx}{eq:gammadFapprx}, but is just a manifestation of a scale invariance. The existence of $\psi$ corrects this perturbatively. Note that $F(f)$ is invariant under the scale transform $x\rightarrow \lambda x$ and $w\rightarrow \lambda^{-2}w$ if there were no lens. Since the symmetry is broken by $\psi$, we keep track of the effects by a spurion coupling $a\psi$ that compensates the symmetry breaking. For simplicity, by considering a power-law profile $\psi \propto x^{2-k}$ (\Sec{sec:discussion}), the scale invariance $w a x^{2-k}\rightarrow w a x^{2-k}$ requires $a\rightarrow \lambda^ k a$. The leading term of the perturbation expansion (in powers of $a$) of $F(f)$ must be of the form 
\begin{align} \label{eq:weakpower}
    F(f) \= a \psi w^q +\text{const.}\, , \qquad \text{with }\ q\=\frac{k}{2}\, 
\end{align}
to respect the scale invariance. The power of $w$ is thus uniquely determined by the spurious scale invariance, and indeed agrees with our power-law calculation in \Sec{sec:power-law}. On the other hand, in the strong diffraction regime, the Einstein radius  fixes the length scale of $F(f)$ (as a stationary point), and a scale invariance no longer exists. Therefore, the existence of a scale invariance discriminates strong/weak diffractive lensing.

The frequency independence of geometric optics is also explained similarly. In this regime of $r_F \lesssim \max( r_s^2, 2\sqrt{r_s r_E})$ and $r_s\neq 0$, only stationary points of $\hat{T}_d(\bm{x})$ (hence, separate images) contribute to \Eq{eq:dimlessF}. In the small neighborhood of each image, a scale invariance holds and, as a result, the contribution of each image to $F(f)$ is constant. If there are multiple images, $F(f)$ also contains the interference between them, which becomes increasingly oscillatory with $w$.

\medskip
As an interesting aside, we can understand the frequency dependencies in yet another way. We can derive them just by matching $F(w)$ to geometric optics at the diffraction boundaries $r_F=r_s$ \Eq{eq:waveoptics2} and $r_F^2 = 4 r_s r_E$ \Eq{eq:generalcond}.
For weak diffraction, matched at $r_F = r_s$, the geometric-optics magnification of the single image at $r_s$ is
\bea
F &\=& \sqrt{\frac{1}{ (1-\kappa(r_s))^2 - \gamma(r_s)^2 }} \, \simeq \, 1+ \kappa(r_s) \\
&\=& 1 + \frac{2-k}{2} x_s^{-k} \= 1 + \frac{2-k}{2} \left(\frac{w}{2}\right)^{\frac{k}{2}}
\eea
where $\kappa, \gamma \ll 1$ and in the second line we have used power-law results derived  in \Sec{sec:discussion}. This indeed has $F-1 \propto w^{k/2}$ as in \Eq{eq:weakpower}. For strong diffraction, matched at $r_F^2 = 4 r_s r_E$, the geometric-optics magnification of one of the multi-images located at $x_i = 1+\delta x =1+ x_s/(1-\psi^{\prime \prime}(1))$ \Eq{eq:deltax} is ($\delta x \ll 1$)
\bea
F &\=& \sqrt{\frac{1}{(1-\kappa(x_i)^2) - \gamma(x_i)^2}}  \,\simeq\, \sqrt{\frac{1}{k x_s}} \= \sqrt{\frac{2w}{k}}.
\eea
where again we have used power-law results.
This indeed has $F \propto w^{1/2}$ as in \Eq{eq:strongdiff}; the dependence of $k^{-1/2}$ is also correct as in \Eq{eq:powerFstrong}. Thus, the physics of the wave-to-geometic optics boundary and geometric-optics magnification already contain the $w$-dependencies.

\medskip
This completes the formalism of diffractive lensing. In the next few sections, we apply the weak diffraction to NFW, while in \Sec{sec:discussion} we apply the full formalism to general power-law profiles.

%%%%%%%%%%
\section{NFW lensing} \label{sec:NFW}

As an important example, we work out diffractive lensing by NFW using our formalism.

%%%%%
\subsection{Profile}

The NFW profile~\cite{Navarro:1995iw} is commonly used to parametrize spherically symmetric density profiles of CDM halos. With two parameters, $\rho_0$ and $r_0$, its three-dimensional radial profile is given by 
\begin{equation}
    \rho(r) \= \frac{4\rho_0}{(r/r_0)(1+r/r_0)^{2}},
\end{equation}
where $r$ is the radial distance from the center, $r_0$ the scale radius at which the slope of profile turns from $-1$ inside to $-3$ outside, and $\rho_0$ the mass density at $r_0$. Since the total mass diverges, this profile must be cut off at some $r$ not far from $r_0$; only the scale $r \lesssim r_0$ will be relevant to the lensing.
The surface density at the distance $x = r/r_0$ from the center on the lens plane is given by~\cite{Keeton:2001ss}
\beq \label{eq:surfaced}
    \Sigma(x) \= \int_{-\infty}^{\infty} dz \, \rho \left(\sqrt{x^2 r_0^2+z^2}\right) 
     \= 3\Sigma_0 \frac{1-\mathcal{F}(x)}{x^2-1},
\eeq
where $\Sigma_0 = 8\rho_0 r_0/3 =\Sigma(x=1)$ and
\begin{equation}
    \mathcal{F}(x)=
    \begin{cases}
       \frac{\text{arctanh}\sqrt{1-x^2}}{\sqrt{1-x^2}} & x<1\\
        1 & x=1\\
        \frac{\text{arctan}\sqrt{x^2-1}}{\sqrt{x^2-1}} & x>1
    \end{cases}\ .
\label{eq:curFNFW} \end{equation}
The 2D Poisson equation \Eq{eq:laplacian} is solved as
\begin{equation} \label{eq:psiNFW}
  \psi(x) \= 3\kappa_0 \left[\ln^2 \frac{x}{2} +(x^2-1)\mathcal{F}^2(x)\right]\, ,
\end{equation}
where $\kappa_0 = \Sigma_0/\Scrit$.

\begin{figure}
\includegraphics[width=0.85\linewidth]{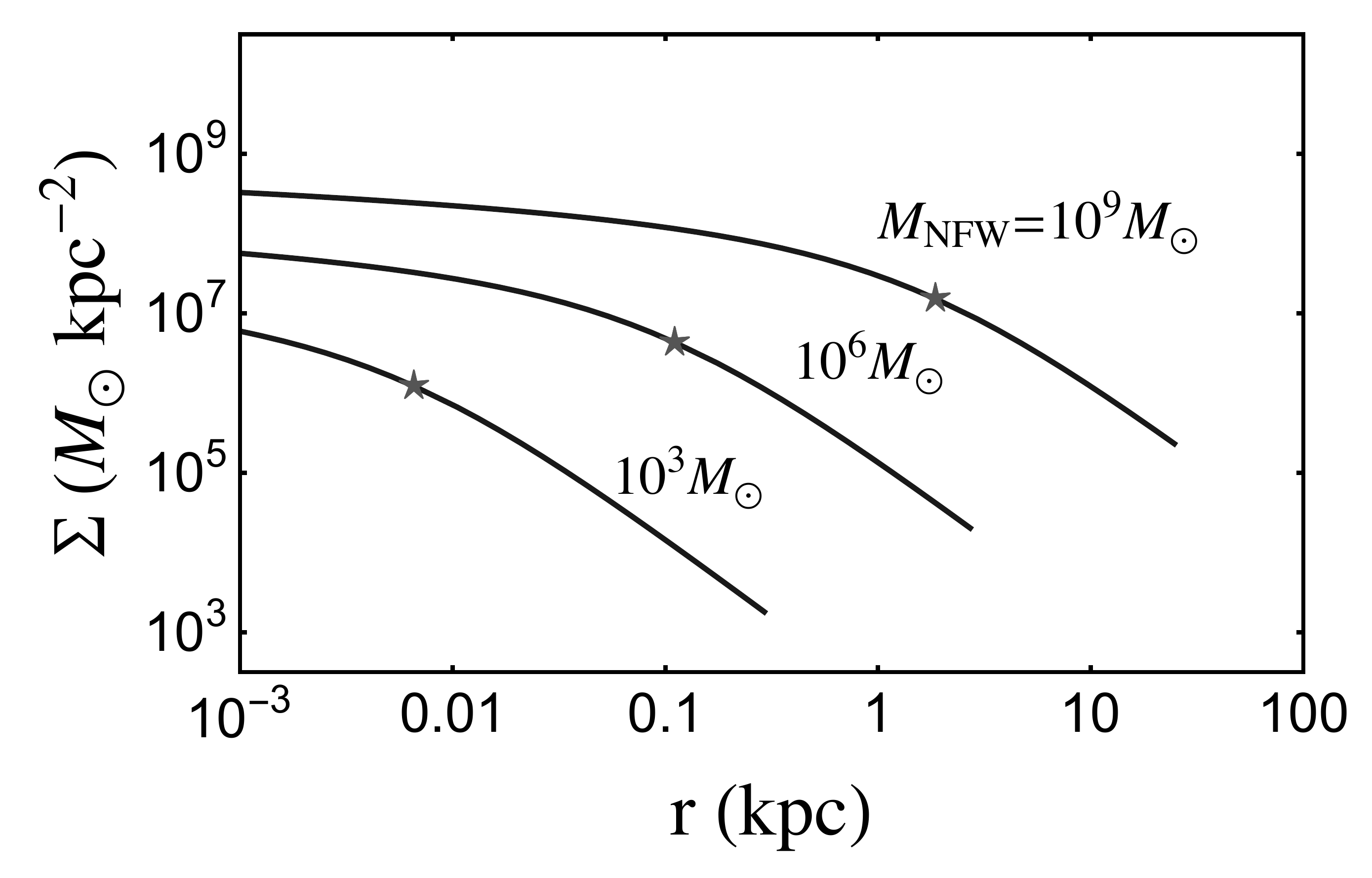}
\caption{\label{fig:surfaced}
The surface mass density $\Sigma(r)$ at the distance $r$ from the center of the NFW profile, with $\MNFW = 10^9, 10^6, 10^3 \Msun$. The star on each curve denotes the location of the scale radius $r_0$. The curves end at their virial radius, $r_\text{vir} = c r_0$, where $c$ is given by the Okoli's $\Mvir-c$ relation~\cite{Okoli:2015dta}. 
} 
\end{figure}

The NFW parametrization is simplified by removing one of the two parameters using the $\Mvir-c$ relation predicted by CDM simulations. Here, $\Mvir$ is the virial mass of a halo, 
and $c \equiv r_{\rm vir}/r_0$ is the concentration. We take the $\Mvir-c$ relation at $z=0$ from Okoli et. al. in~\cite{Okoli:2015dta}. 
%We do not consider its redshift evolution, for simplicity. 
Moreover, instead of conventional $M_\text{vir}$, it is more convenient to use the NFW mass defined as
\beq
    \MNFW \,\equiv \, 16\pi \rho_0 r_0^3\,
\eeq
because it represents the halo mass independently of redshift. The two masses are related by
\beq
    \Mvir \= M_\text{NFW} \left(\, \ln(1+c)-c(1+c)^{-1} \, \right),
\eeq
differing only by ${\cal O}(1)$ as $c = 10 \sim 50$ for $\Mvir= 10^4 \sim 10^{10} \Msun$~\cite{Okoli:2015dta,Okoli:2017uts}.

Now, $\MNFW$ fixes all the parameters of the NFW profile. For example, we can express most relevant lens properties in terms of $\MNFW$ as (using central values of the Okoli's relation)
\bea \label{eq:CDMMS}
    \Sigma_0 &=& \frac{8}{3} \rho_0 r_0 \,\simeq\, 1.3 \times 10^{7}\Msun/\text{kpc}^2 \left( \frac{\MNFW }{ 10^9\Msun} \right)^{0.18}, \\
    r_0 &=& \sqrt{\frac{\MNFW}{6 \pi \Sigma_0}} \, \simeq \, 2\ \text{kpc} \left( \frac{\MNFW }{ 10^9 \Msun} \right)^{0.41}.
\label{eq:rsMNFW}
\eea
Fig. \ref{fig:surfaced} shows the surface mass density $\Sigma(x)$ and $r_0$ for $\MNFW=10^3, 10^6,$ and $10^9\Msun$. $\Sigma(r)$ is obviously smaller for lighter halos while not varying rapidly inside $r_0$; thus, $\Sigma_0 = \Sigma(r=r_0)$ or $\kappa_0$ characterizes the values of $\Sigma(x)$ or $\kappa(x)$. $r_0$ is smaller for lighter NFWs, and it is the length scale relevant to this work. We collect other useful expressions too
\bea \label{eq:kappa0}
    \kappa_0 \,&=&\, \frac{\Sigma_0}{\Scrit} \, \simeq \, 7.9 \times 10^{-3} \left( \frac{\MNFW}{10^9 \Msun} \right)^{0.18} \left( \frac{d_\text{eff} }{\rm Gpc } \right), \\
\Scrit \,&=&\, \frac{1}{4\pi d_{\rm eff}}  \, \simeq \, 1.66\times10^9\ \Msun \text{kpc}^{-2}\left(\frac{\rm Gpc }{d_{\rm eff}}\right).
\eea

\begin{figure*}
\includegraphics[width=0.45\linewidth]{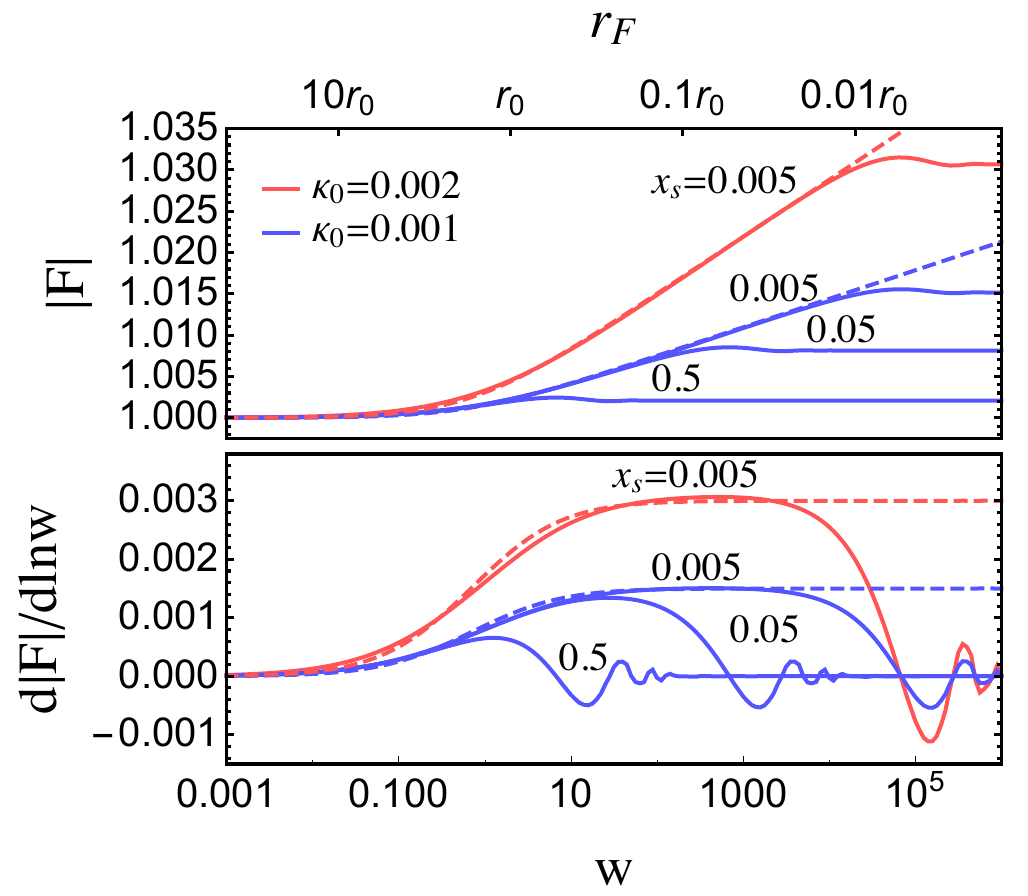}
\includegraphics[width=0.45\linewidth]{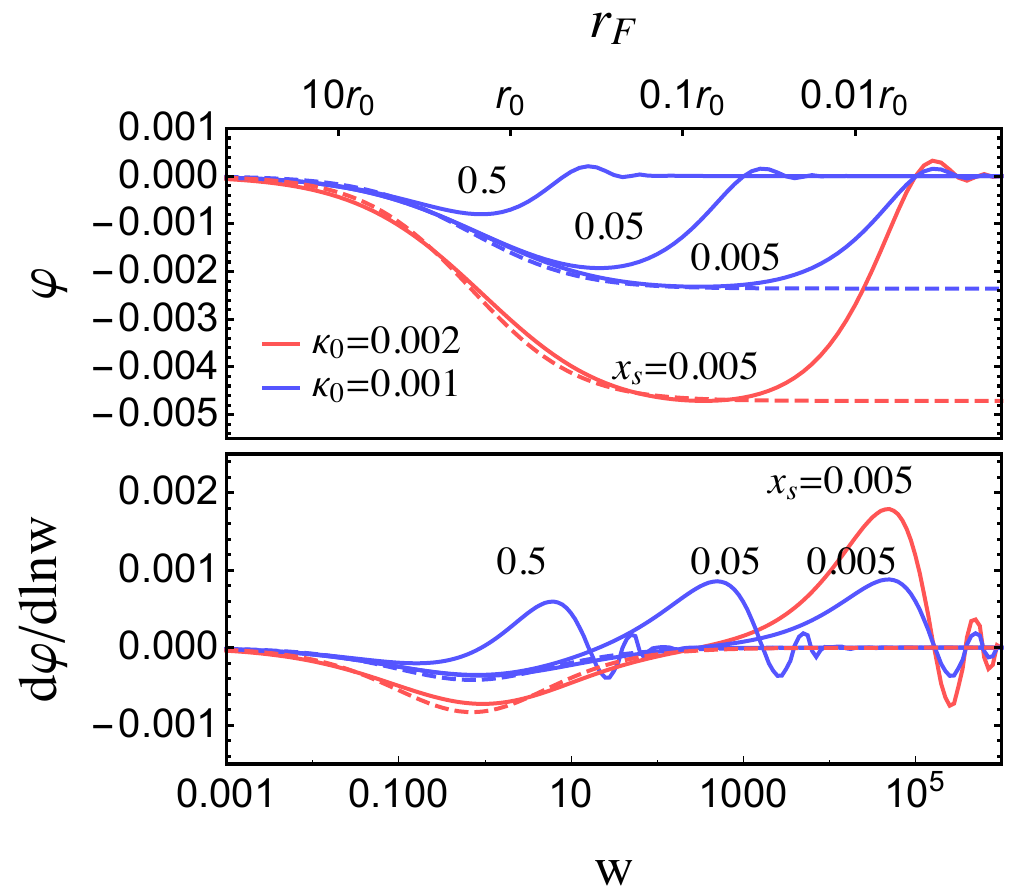}
\caption{ \label{fig:FdFex}
$|F(w)|$ (upper left), $\frac{d|F(w)|}{d \ln w}$ (lower left), $\varphi(w)$ (upper right), and $\frac{d\varphi(w)}{d \ln w}$ (lower right) for NFW profiles with $\kappa_0=0.002$ (red) and $\kappa_0=0.001$ (blue). Solid lines are full numerical solutions of \Eq{eq:Fshift}, while dashed are diffraction-limit results in \Eqs{eq:NFWFapprx}{eq:NFWdFapprx}. Their $|F(w)|$ and $\varphi(w)$ are obtained according to Eqs. (\ref{eq:Fabs}$\sim$\ref{eq:dFphase}). All of them agree in the diffraction regime $w  \lesssim 2/x_s^2$; see more in text. Each curve is marked with $x_s$ value. 
} 
\end{figure*}
%

%%%%%
\subsection{Critical curves} \label{sec:critical}

%The critical curves of NFWs are extremely small so that multi-imaged lensing is very rare. 
Critical curves are the locations of images where their magnifications (formally) diverge. The magnification in the geometric-optics limit
\bea
    \mu &=&  [ \, \det A(\bm{x}) \, ]^{-1} \label{eq:mudet} \\
    &=& \left[ (1-\kappa)^2 -\gamma^2 \right]^{-1} \= \left[ \left( 1- \frac{\psi^\prime }{ x} \right) \left(1 - \psi^{\prime \prime} \right) \right]^{-1},
%    \left( \left(1-x^{-1}\frac{d}{dx}\psi_\text{NFW}\right)\left(1-\frac{d^2}{dx^2}\psi_\text{\NFW}\right)\right]^{-1}
\eea
where $A(\bm{x}) \equiv d \bm{x}_s / d\bm{x}$ is a $2\times 2$ matrix of the $\hat{T}_d$ curvature around the image, yields two such solutions 
\beq \label{eq:tangentialC}
    x_t \,\simeq\, 2\exp\left[-\tfrac{1}{2}-\tfrac{1}{3\kappa_0}\right], \,\,\, x_r \,\simeq\, 2\exp\left[-\tfrac{3}{2}-\tfrac{1}{3\kappa_0}\right],
\eeq
called tangential and radial critical curves, respectively.
$x_t$ is also called the Einstein radius $x_E$. 
Since $\kappa_0 \lesssim 10^{-2}$ is small for NFWs considered in this work, the critical curves are exponentially suppressed $x_{t,r}\lesssim \exp(-100) \ll 1$ and $x_E$ is essentially zero. What does this mean?

Critical curves (more precisely, caustics) are roughly the boundary between regions of different number of images; if $\det A [$ \Eq{eq:mudet}] does not change its sign, then the mapping between source $\bm{x}_s$ and image $\bm{x}$ planes is one-to-one invertible so that there can only be a single image~\cite{Schneider:1992}.
Also, critical curves are (more precisely, Einstein radius) the boundary between geometric versus Shapiro time delay dominance. 
Therefore, NFW lensing is always single imaged (see also \App{app:singleNFW}) and governed by geometric time delay; leading gravitational effects come from the perturbation of order $\psi$ near the image. But this is not a general property of diffuse profiles as will be discussed in \Sec{sec:power-law}.

How can single-imaged lensing be detected?
Again, it is possible using the frequency dependence of diffractive lensing and the frequency chirping of GW.

%%%%%
\subsection{Diffractive lensing} \label{sec:NFWlensing}

We solve NFW (weak) diffractive lensing analytically.
Plugging \Eq{eq:psiNFW} into \Eqs{eq:aximeanconv}{eq:aximeanshear}, we have
\begin{align}
    \overline{\kappa}(x)&\= \frac{6\kappa_0}{x^2}\left[\ln \frac{x}{2} + \mathcal{F}(x)\right], \label{eq:NFWmeanconv}\\
   \gamma(x)&\=\frac{6\kappa_0}{x^2}\left[\ln\frac{x}{2}+\mathcal{F}(x)-\frac{x^2}{2}\frac{1-\mathcal{F}(x)}{x^2-1}\right],\, \label{eq:NFWshear}
\end{align}
where $\mathcal{F}(x)$ is given in \Eq{eq:curFNFW}. Then, according to \Eqs{eq:meankFapprx}{eq:gammadFapprx}, the analytic continuation of \Eqs{eq:NFWmeanconv}{eq:NFWshear} yields
\beq \label{eq:NFWFapprx}
F(w) \,\simeq\, 1-6\kappa_0  i w \left[ \frac{i \pi  }{4}-\frac{1}{2}\ln w -\ln 2 + \mathcal{F}(w^{-\frac{1}{2}}e^{\frac{i\pi }{4}})\right],
\eeq
\bea 
\frac{dF(w)}{d \ln w} &\simeq& -6\kappa_0  i w \left[ \frac{i \pi  }{4}-\frac{1}{2}\ln w -\ln 2 + \mathcal{F}(w^{-\frac{1}{2}}e^{\frac{i\pi }{4}}) \right. \nonumber\\
&& \qquad \qquad \left. -\frac{i}{2 }\frac{1-\mathcal{F}(w^{-\frac{1}{2}}e^{\frac{i \pi }{4}})}{i-w}\right]\, .
\label{eq:NFWdFapprx} \eea
Recall that this derivation is valid for $r_F \gtrsim r_s$ and $r_s \gtrsim r_E$, but since $r_E$ vanishes for NFW these results are valid for all $w=2(r_0/r_F)^2$ as long as $r_F \gtrsim r_s$. Although these are complicated functions of $w$ in general, they are simplified in the limits of $w \gg 1$ and $\ll 1$. For $w\ll 1$ ($w\gg 1$), they asymptote as $dF/d \ln w  \propto w$ ($\propto $ const), which agrees with the results of $k\rightarrow 2$ ($k\rightarrow 0$) power-law profiles since these limits correspond to the outer (inner) part of NFW with $\rho \propto r^{-3} (r^{-1})$.

%\medskip
\Fig{fig:FdFex} above all, confirms these analytic solutions (dashed)  in the diffractive regime of $w \lesssim 2/x_s^2$, compared with full numerical results of \Eq{eq:physicalF} (solid). Around this boundary, they are matched well to the well-known geometric-optics results. Therefore, it is remarkable that one can understand the results of a complicated lensing integral in terms of much simpler 2D potentials.
 
\Fig{fig:FdFex} further demonstrates the main features of NFW diffractive lensing. In the diffraction regime, both amplification $|F(w)|$ and phase $\varphi(w)$ are frequency dependent, as expected. Its strength does not depend on $x_s$ (i.e., $x_s$ not resolved) so that blue curves with different $x_s$ coincide there. But $x_s$ determines at which frequency lensing becomes geometric optics (i.e., when $x_s$ is resolved). As a result, larger lensing effects can be obtained for smaller $x_s$; geometric-optics lensing is stronger for sources closer to the lens. Soon after geometric optics is reached, the slopes of $|F(w)|$ and $\varphi(w)$ vanish, and lensing becomes frequency independent. Lastly, single-imaged diffraction always amplifies the wave, as also proved in \App{app:singleNFW}. 

Notably, $\varphi(w)$ itself also vanishes in the geometric-optics limit. It is because $T_0$ was factored out in \Eq{eq:Fshift} so that the single image in this limit does not have extra phases; we will see $\varphi(w)$ for multi-imaged cases in \Sec{sec:discussion}. Although the frequency dependence of $\varphi(w)$ is more complicated than that of $|F(w)|$, their overall sizes are similar, commonly given by $\overline{\kappa}(x)$ and $\gamma(x)$.

%%%%%%%%%%%
\section{GW detection of NFW} \label{sec:detection}

We introduce the concept of detection with chirping GW and likelihood criteria for detection.

%%%%%
\subsection{GW chirping} \label{sec:GW}

One of the most important features of GW is that its amplitude and frequency ``chirp''. It is worth emphasizing that what is actually measurable is the frequency-dependent change of lensing effects, not the absolute size of amplification.

The observed unlensed chirping amplitude in the frequency domain can be written as
\begin{align}
    \tilde{h}(f) \= A_p A(f) e^{i (2\pi f t_c^0 + \phi_c^0 + \Psi(f))}.
\label{eq:waveform0} \end{align}
The chirping $A(f)$ with particular frequency dependencies as described below will be the basis of lensing detection, while the chirping phase $\Psi(f)$ will be canceled out between lensed and unlensed waveforms [see \Eq{eq:lnp}].
Coalescence time $t_c^0$ and constant phase $\phi_c^0$ set to zero for the best-fit procedure[see \Eq{eq:bestfit}] since they can be arbitrary.
For simplicity we fix binary and detector parameters (polarization, binary inclination, and detector antenna direction) such that $A_p=1$, and ignore black hole spins and detector reorientation during measurements; such effects will in principle be distinguishable from lensing effects. We refer to \cite{Jung:2017flg,Dai:2018enj} for more discussions on this simplified analysis.

\begin{figure}[t]
\includegraphics[width=0.85\linewidth]{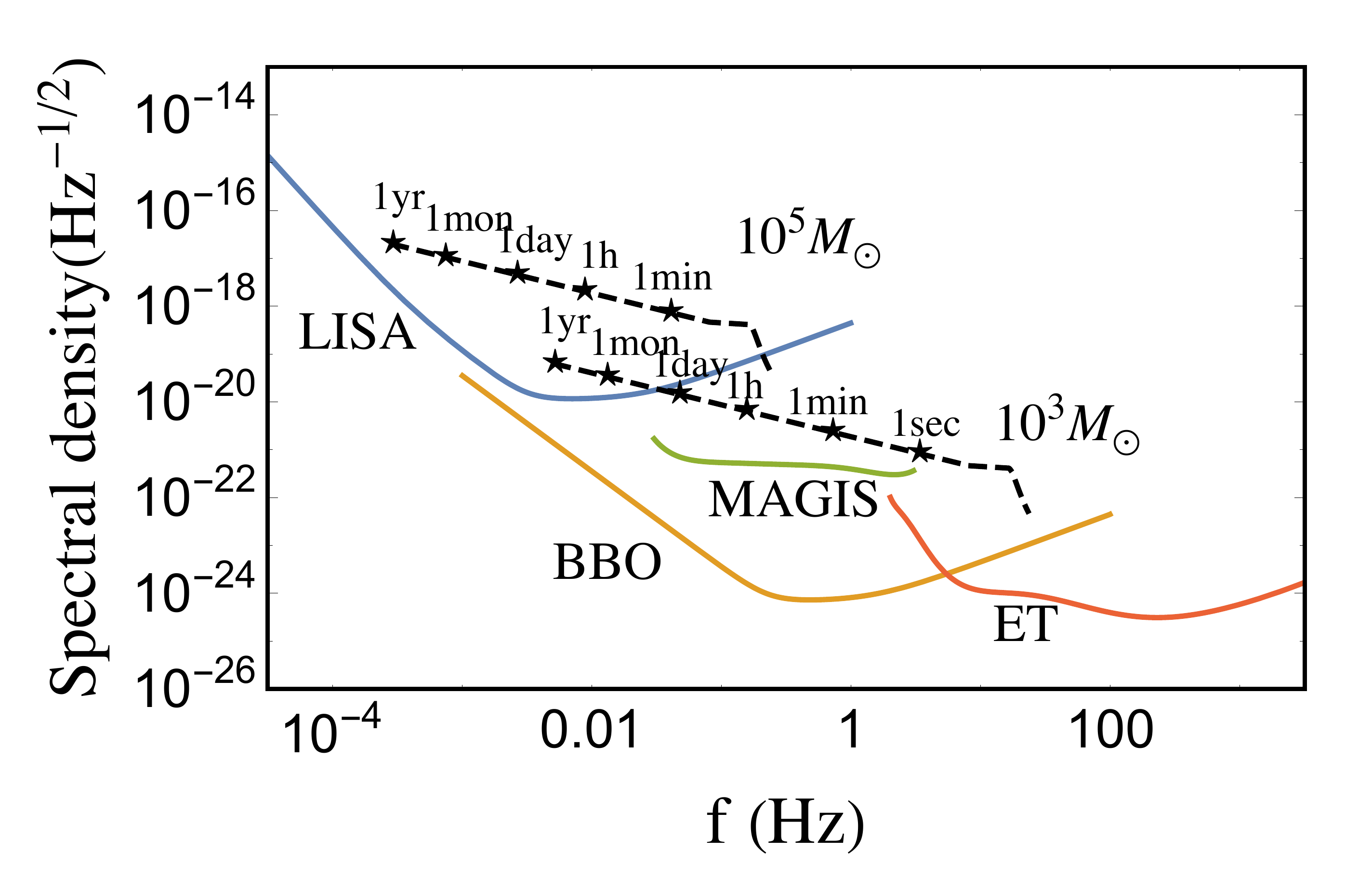}
\caption{ \label{fig:sensitivity}
The spectral density of GW detector noises $\sqrt{S_n(f)}$ (solid) and example chirping GW amplitudes $\sqrt{f}|\tilde{h}(f)|$ (dashed) with $\MBBH =10^3\, \Msun$ and $10^5\, \Msun$. $z_s=1$. The time marked with stars indicate the remaining time until final merger.} 
\end{figure}

The frequency dependence of $A(f)$ differs in the successive phases of  inspiral-merger-ringdown. For the inspiral phase $f<f_\text{merg}$, we adopt PhenomA waveform templates developed in Ref. \cite{Ajith:2007kx}, approximating nonspinning quasicircular binaries. The waveform is
\begin{align}
    A(f)=A_\text{insp}(f) =\sqrt{\frac{5}{24}} \frac{\mathcal{M}^{\frac{5}{6}}f^{-\frac{7}{6}}}{\pi^{\frac{2}{3}}d_L}\, ,
\end{align}
which is the restricted post-Newtonian approximation.
The chirp mass $\mathcal{M}= \MBBH / 2^{6/5}$ for equal-mass binaries with the total mass $\MBBH$ (we consider only such cases), and $d_L$ is the luminosity distance to the source. All the masses are redshifted ones.
The amplitude in the merger ($f_\text{merg}\leq f<f_\text{ring}$) and ringdown phases ($f_\text{ring}\leq f<f_\text{cut}$) are
\begin{align}
    A(f)=A_\text{insp}(f_\text{merg})\times
    \begin{cases}
    \left(\frac{f}{f_\text{merg}}\right)^{-2/3}& \text{merger}\\[10pt]
    \frac{\sigma_f^2/4}{(f-f_\text{ring})^2+\sigma_f^2/4}& \text{ringdown}
    \end{cases}\, ,
\end{align}
where $\sigma_f$ is the width of a peak centered at $f_\text{ring}$. 
 The expressions for $f_\text{merg}$, $f_\text{ring}$, $f_\text{cut}$, and $\sigma_f$  are detailed in Ref. \cite{Ajith:2007kx}. 
Example chirping waveforms $|\tilde{h}(f)|$ based on these expressions are shown in \Fig{fig:sensitivity}. Frequency-dependent lensing effects will be detectable as a deviation to the chirping.

Also marked on the chirping waveforms are the time remaining until final merger. The frequency chirping in time at leading post-Newtonian order is given by 
\begin{align}
    f(t) &= \frac{1}{8\pi \mathcal{M}}\left(\frac{5\mathcal{M}}{t}\right)^{3/8}= 0.39\ \text{Hz} \left(\frac{\Msun}{\MBBH}\right)^{5/8}\left(\frac{\text{yr}}{t}\right)^{3/8}
\end{align}
for time $t$ before final merger. Almost all of the time is spent during the inspiral.

The benchmark GW detectors are Laser Interferometer Space Antenna (LISA) \cite{Audley:2017drz,Cornish:2018dyw}, Big Bang Observer (BBO) \cite{Cutler:2009qv}, Matter-wave Atomic Gradiometer Interferometric Sensor (MAGIS) \cite{Graham:2017pmn,Graham:2016plp}, and Einstein Telescope (ET) \cite{Punturo:2010zz}.
%All of them except ET are space-based detectors. LISA, which covers lowest frequency range among them, is designed to sensitive to supermassive binary black hole mergers($M\sim 10^6\, \Msun$). We calculate the power spectral density of its noise following \cite{Cornish:2018dyw}, but ignore galaxy background. BBO produces the largest Signal-to-Noise Ratio(SNR) among all the detectors but sensitive to lighter masses $M\sim 10^5\, \Msun$ compared to LISA. 
%MAGIS, assuming it is operating in the resonant mode\cite{Graham:2016plp}, is also sensitive to $M\sim 10^5\, \Msun$, but the SNR produced by the binary merger is smaller than that of BBO and similar to that of LISA. 
%ET is operating at highest frequency range and sensitive to $M\sim 10^3\, \Msun$. 
Their noise spectral densities $S_n(f)$ are shown in \Fig{fig:sensitivity}. The sensitivity ranges are roughly $[10\, \mu\text{Hz}, 1 \, \text{Hz}]$ (LISA), $[1\, \text{mHz}, 100 \, \text{Hz}]$ (BBO), $[30\, \text{mHz}, 3\, \text{Hz}]$ (MAGIS), and $[2\, \text{Hz}, 10 \, \text{kHz}]$ (ET).

%%%%%
\subsection{Log-likelihood detection}  \label{sec:criteria}

How well can the single-imaged diffractive lensing be detected? Detection likelihood is measured by~\cite{Jung:2017flg,Dai:2018enj} 
\beq
\ln p \=  -\frac{1}{2}   ( h_L - h_{\rm BF} | h_L - h_{\rm BF} ),
\label{eq:lnp} \eeq
where $h_{\rm BF}$ is the best-fit `unlensed' GW waveform that maximizes the likelihood. 
The best-fit is performed with respect to the overall amplitude $A$, constant phase $\phi_c$, and coalescence time $t_c$ of the unlensed $\tilde{h}(f)$ \Eq{eq:waveform0} as 
\begin{align}\label{eq:bestfit}
%    \tilde{h}_L(f) &= F(f) \tilde{h}_0(f)\, ,\\
    \tilde{h}_{\rm template}(f) &=  \tilde{h}(f) A e^{i(2\pi f t_c +\phi_c)}.
\end{align}
When $h_{\rm BF}$ perfectly matches $h_L$, $A=1$, and $t_c=\phi_c=0$.
The inner product $(h_1|h_2)=  4 \text{Re} \int df \, \tilde{h}^{*}_{1}(f)\tilde{h}_{2}(f)/S_n(f)$, where $S_n(f)$ is the noise spectral density. The best-fit in this way is discussed more in \cite{Jung:2017flg,Dai:2018enj}.

In this way, the $\ln p$ measures how well lensed signals can be fitted with unlensed waveforms. 
%Non-trivial frequency dependent lensing effects will not be fitted by a constant $A$, among others. 
Frequency dependent lensing amplitude $|F(f)|$ will not be fitted by a constant $A$. 
Likewise, nontrivial frequency dependent lensing phase $\varphi(f)$ cannot be canceled by $\phi_c$ and $t_c$. 
Thus, the larger the $|\ln p|$, the worse the best fit, hence the more confident is the existence of lensing. 

In principle, the larger $|\ln p|$, which is equivalent to the smaller match,\footnote{$(\text{match})\equiv( h_L| h_{\rm BF} )/\sqrt{( h_L| h_L ) ( h_{\rm BF}| h_{\rm BF} )}$}
 reduces the ability of GW detection. However, we can ignore such effects in our NFW lensing situations thanks to the small mismatch($1-(\text{match})\simeq 10^{-6}$).
\footnote{Actually, from \Eq{eq:lnpapprx}, one can easily show that the mismatch is approximately given by the square of the shear of the lens object.}
In spite of the small mismatch, the lensed GW can be distinguished from the unlensed GW if the SNR of the GW waveform is sufficiently high\cite{Dai:2018enj,Mishra:2021xzz}.

In this work, the binary intrinsic parameters like total mass, mass ratio, and spins are not included in the best-fit procedure. We expect that taking into account the binary parameters will not significantly reduce $|\ln p|$ values. This is because, the frequency dependence of $F(f)$ around the diffraction-geomtric optics transition frequency(e.g. \Fig{fig:FdFex}) is characteristically different from the intrisic frequency dependence of GW waveform even if post-Newtonian corrections are considered.
More accurate analysis on the potential degeneracy between the diffractive lensing and the GW waveform are beyond our scope and should be explored in future research. 

\begin{figure}
\includegraphics[width=0.8\linewidth]{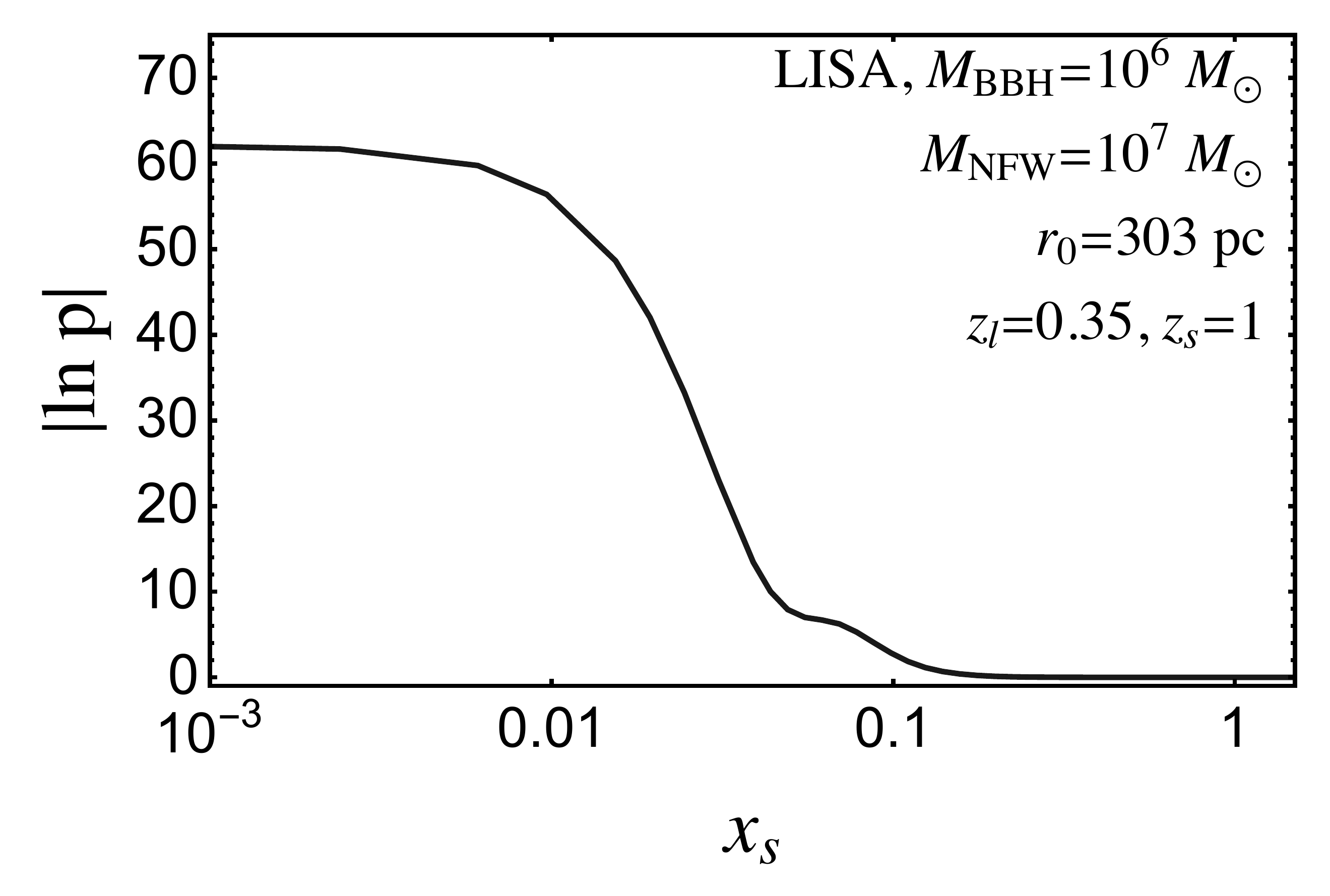}
\caption{ \label{fig:minlogp}
$|\ln p|$ as a function of $x_s=r_s/r_0$. Last one year of inspiral observed at LISA.} 
\end{figure}

We require $\ln p < -5.914$ for $3\sigma$ confidence of the lensing detection. The requirement yields a proper lensing cross section for given masses and distances
\beq
\sigma_l \= \pi (r_0 x_s^{\rm max})^2.
\label{eq:sigNFW} \eeq
There exists a maximum $x_s^{\rm max}$ for given parameters because $|\ln p|$ generally decreases with $x_s$ as shown in \Fig{fig:minlogp}. If there exist multiple roots of $x_s^{\rm max}$, we take the largest one, while if no root $x_s^\text{max}=0$. An example result of $x_s^\text{max}$ is shown in \Fig{fig:sigmacontour} in \App{app:lencrx}. In later sections, $\sigma_l$ will be used for lensing probabilities.

\medskip
For numerical calculation, a more convenient form for $\ln p$ is obtained by analytically minimizing $\ln p$ with respect to $A$ and $\phi_c$ as
\begin{align}
    \ln p \= -\frac{1}{2}(\rho_L^2 - \rho_{uL}^2)\, ,
\end{align}
where 
\bea
    \rho_L^2 &=& (h_L|h_L)\, ,\\
    \rho_{uL}^2 &=&  \max_{t_c} \left|\frac{4}{\rho_0}\int_{f_\text{min}}^{f_\text{max}}df \frac{ |\tilde{h}_0(f) |^2 }{S_n(f)}F^*(f)e^{2\pi i f  t_c}\right|^2\,  \label{eq:rhouL}
\eea
and $\rho_0^2=(h_0|h_0)$ is SNR squared. Here, the maximization with respect to $t_c$ should be done numerically; but $t_c$ maximization is relatively unimportant since adding $T_0$ in \Eq{eq:Fshift} approximately does this maximization. More discussions are presented in Ref.~\cite{Dai:2018enj} and in \App{app:formula-lnp}.

As an aside, there also exists the maximum $|\ln p|$ for some small $x_s$ for given lensing parameters. As shown in \Fig{fig:minlogp}, $|\ln p|$ stops growing for $x_s \lesssim 10^{-2}$. It is because, for small enough $x_s$, diffraction occurs in the whole frequency range of measurement so that diffraction amplification does not depend on $x_s$ as shown in \Fig{fig:FdFex}. Under this condition, we find that 
\begin{align}\label{eq:lnpapprx}
    |\ln p| \,\simeq \,
    \frac{1}{8}\left\{ \rho_0 \cdot \left| \gamma\left( \frac{r_F(f_0)e^{i\frac{\pi}{4}}}{\sqrt{2}}\right) \right| \cdot \ln \frac{f_\text{max}}{f_\text{min}} \right\}^2\, ,
\end{align}
where $f_0$ is a characteristic frequency at which 
\begin{align}\label{eq:f0def}
    \frac{\rho_0^2}{2}=4\int_{f_0}^{f_\text{max}}df \frac{|h_0(f)|^2}{S_n(f)}.
\end{align}
$f_0$ is typically close to the maximum point of $|h_0(f)|^2/S_n(f)$. Its derivation is given in \App{app:maxlnp}. \Eq{eq:lnpapprx} also supports our intuition that the strength of shear is critical to lensing detection.

%%%%%%%%%%%
\section{Prospects} \label{sec:results}

We first develop intuitions by semianalytically estimating the parameter space of NFW lensing, and then obtain final results with full numerical calculation.

%%%%%
\subsection{Semianalytic estimation} \label{sec:MNFW}
%%%%

Which NFW mass scale is relevant to diffractive lensing? Since diffractive lensing is sensitive to the mass profile at $r_F$ through shear $\gamma(r_F)$ \Eq{eq:gammadFapprx}, the profile must have sizable shear in the chirping range of $r_F$. For NFW, this happens if some range of $r_F$ satisfies
\beq \label{eq:rsrange}
 10^{-3}r_0 \,\lesssim \, r_F  \, \lesssim \, r_0.
\eeq
The maximum is restricted to be within $r_0$ because it is where $\gamma \sim 3 \kappa_0/2$ is most sizable; outside, gravity is suppressed quickly with $\gamma \propto 1/x^2$. The minimum $10^{-3} r_0$ is introduced for the ease of calculation and is chosen arbitrarily; the area within the minimum is small enough not to affect lensing probability, and the inner profile may be uncertain too.
Therefore, the relevant $\MNFW$ is the one whose length scale $r_0$ is comparable to the range of $r_F$.

\begin{figure}
\includegraphics[width=\linewidth]{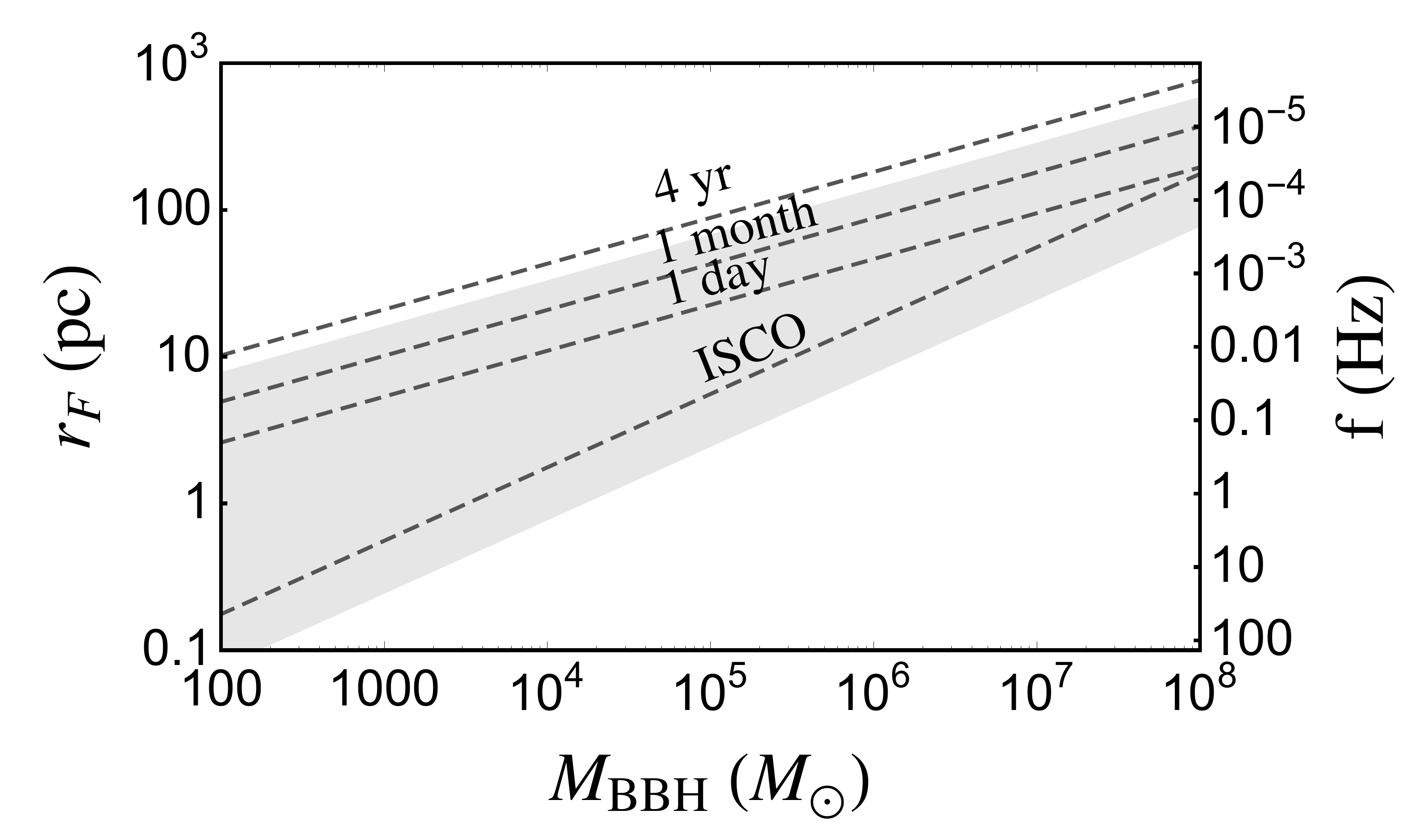}
\caption{ \label{fig:rFrange}
The range of Fresnel length \Eq{eq:rF} swept by a chirping GW during its last one year before merger (shaded). Other time periods are shown as dashed lines; ISCO refers to the innermost stable circular orbit.  The corresponding GW frequencies are shown on the right vertical axis. Some part of this range, combined with detector sensitivities, must satisfy \Eq{eq:rsrange} for diffractive lensing. $z_s = 1, z_l = 0.35$. 
}
\end{figure}

The chirping range of $r_F \propto f^{-1/2}$ (hence the range of GW frequency) is determined by the total mass of a binary black hole, $\MBBH$, according to the standard GW chirping; see \Sec{sec:GW}. \Fig{fig:rFrange} shows an example range of $r_F$ swept during the last one year of chirping, as a function of $\MBBH$. Basically, the heavier they are the earlier at lower frequencies they merge. The range spans one or two orders of magnitudes, while not significantly broadened by longer measurements since binary inspiral is much slower when far away from merger. We use the last one-year measurements for numerical results. 

\begin{figure}
\includegraphics[width=0.95\linewidth]{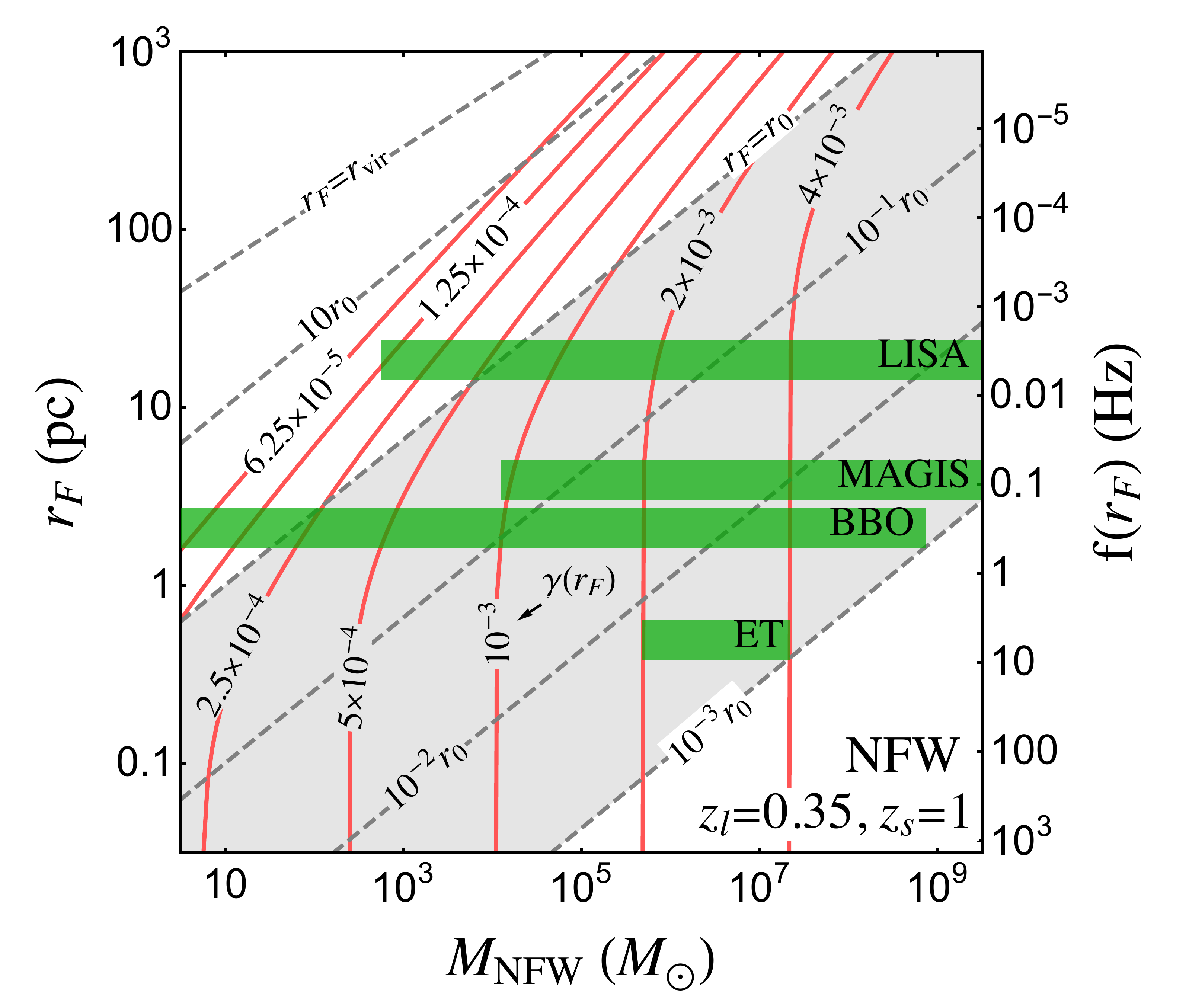}
\caption{ \label{fig:Detectable2}
Semianalytic estimation of the parameter space of NFW diffractive lensing. Diffractive lensing is relevant in the shaded region \Eq{eq:rsrange}. Solid contours show the shear $\gamma(r_F)$, reflecting the required SNR for detection. The frequency corresponding to $r_F$ is shown on the right vertical axis. For quick references, green bars roughly show maximal sensitivities at best frequencies. See text for details. $z_s=1, z_l =0.35$.
}
\end{figure}

\Fig{fig:Detectable2} shows the relevant parameter space of NFW. The shaded region satisfies \Eq{eq:rsrange}, which can be rewritten in terms of $\MNFW$ and $f$ as (using $r_0$ in \Eq{eq:rsMNFW} and $r_F$ in \Eq{eq:rF})
\begin{align}\label{eq:MNFWrange}
    13.6\, \Msun \left(\frac{\rm Hz}{f_{\rm max}}\right)^{1.22}  \lesssim \MNFW  \lesssim 2.82\times10^8\, \Msun \left(\frac{\rm Hz}{f_{\rm min}}\right)^{1.22}.
\end{align}
However, not all this region can be probed; signals must be strong enough. The overall change of amplification --- the detectable signal --- within a modest range of $f$ is $\Delta |F| \sim \gamma(r_F(f_*)) \cdot {\cal O}(1)$  from \Eq{eq:gammadFapprx}, with a characteristic frequency $f_*$ within \Eq{eq:rsrange}. Since the shear of NFW does not vary much within $r_0$ as shown in \Fig{fig:Detectable2}, $\gamma(r_F(f_*)) \sim \gamma(r_0)$. Thus, roughly, 
\beq
{\rm SNR} \, \gtrsim \, 1/\gamma(r_0) \cdot {\cal O}(1)
\eeq
 is needed to detect the diffractive lensing by $\MNFW$. This is somewhat more rigorously justified from \Eq{eq:lnpapprx} and \Fig{fig:lnppower}.
 The contours of $\gamma(r_F)$, reflecting the required SNR, are shown as solid lines.

Based on these, one can now estimate the sensitivity range of $\MNFW$. As quick examples, we show a green bar for each detector, with their maximum SNR at the corresponding frequency; SNR $\simeq 5000,\, 10^5,\, 1000,\, 500$, at $f \simeq 0.004,\, 0.3,\, 0.08,\, 6$ Hz for LISA, BBO, MAGIS, ET, respectively. They roughly show maximal sensitivities, only as quick references. One can see that $\MNFW \lesssim 10^7 \Msun$ is potentially sensitive to all detectors. 
The sensitivity range is indeed estimated by the comparison of the $r_F$ range and the lens scale $r_0$. The lower $\MNFW$ range is limited by too low frequency for LISA and BBO that prohibits diffractive lensing by small $\MNFW$ (SNRs are large enough), or by too small SNR  for MAGIS and ET that prohibits detection of small diffraction. 
Another to note is that, for given $\MNFW$, larger SNR is needed for lower-frequency detectors because corresponding larger $r_F$ probes only outer parts of NFW with smaller shear.

A caveat is that this kind of estimation does not show any lensing probabilities.
In the next subsection, we obtain final results with full numerical calculation, showing lensing probabilities as well as confirming these estimations.

%%%%%%
\subsection{Results}

\begin{figure*}
\includegraphics[width=0.92\textwidth]{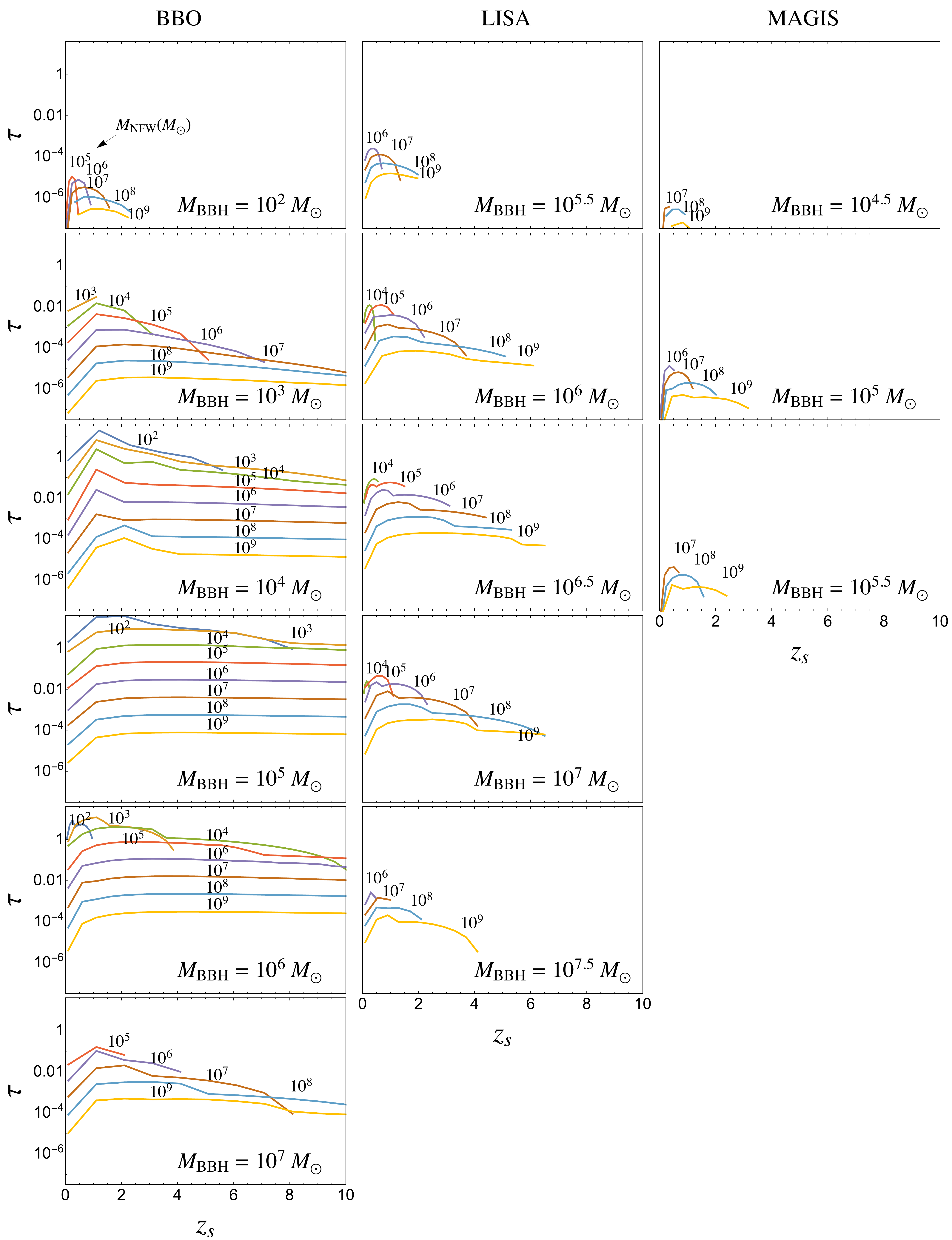}
\caption{\label{fig:tau}
Optical depth $\tau$ for the given $\MNFW$ comprising the full DM abundance (i.e., no halo mass function) at BBO (left), LISA (mid), and MAGIS (right). Each curve is marked with $\MNFW$, and each panel with $\MBBH$. Last one year of inspiral and $3\sigma$ log-likelihood lensing detection.} 
\end{figure*}

We calculate detection prospects, starting from the optical depth (lensing probability).
For the given $\MBBH$, $z_s$ ,and $\MNFW$, the optical depth of the lensing is given by
\begin{equation}
    \tau(z_s)
    \= \int_0^{z_s} d z_l\  \sigma_l (z_l, z_s) \frac{1}{H_0} \frac{ n_l (1+z_l)^2}{\sqrt{(1+z_l)^3 \Omega_m+\Omega_\Lambda}}\ ,
\end{equation}
where $\sigma_l$ is the proper cross section defined in \Eq{eq:sigNFW}.
The comoving DM number density $n_l \=\frac{f_{\rm DM} \Omega_{\rm DM}}{\MNFW} \frac{3H_0^2}{8\pi G}$ is assumed to be constant in $z_l$, with the fraction of mass density $f_{\rm DM}$ to the total DM abundance $\Omega_{\rm DM} =0.25$. Hubble constant $H_0 =70 \,\text{km/s/Mpc}$, and energy density $\Omega_m=0.3$, $\Omega_\Lambda=0.7$ of matter and vacuum energy in units of critical density $\rho_c=3H_0^2/8\pi G$. The lensing probability is $P(\tau) = 1-e^{-\tau} \simeq \tau$ for $\tau \ll 1$.

\Fig{fig:tau} shows the optical depths at LISA, BBO, and MAGIS, for the given $\MNFW$ comprising the total $\Omega_{\rm DM}$(i.e. $f_{\rm DM}=1$ regardless of $\MNFW$); the optical depth at ET is too small to show. Overall, BBO and LISA have sizable $\tau$ close to or even larger than 1, while MAGIS has much smaller $\tau$ at most $\sim 10^{-5}$. This result for single $\MNFW$ can be combined with any mass functions such as given in Refs. \cite{Press:1973iz,Sheth:1999mn}.

The GW lensing event rate $\dot{N}_L$ is obtained by integrating the lensing probability $P(\tau)$ with the comoving merger-rate density $\dot{n}_s$
\beq
\dot{N}_L
\= \int_0^{z_h} dz_s \frac{1}{H_0} \frac{4\pi \chi^2(z_s) }{\sqrt{(1+z_s)^3 \Omega_m+\Omega_\Lambda}} \, \frac{\dot{n}_s}{1+z_s} \, P(\tau), 
\eeq
where $z_h$ is the horizon distance of a GW detector and $\chi(z)$ is the comoving distance. The extra factor $1/(1+z_s)$ accounts for the redshift of the source-frame time period used to define the merger-rate.

\begin{figure}
\includegraphics[width=0.9\linewidth]{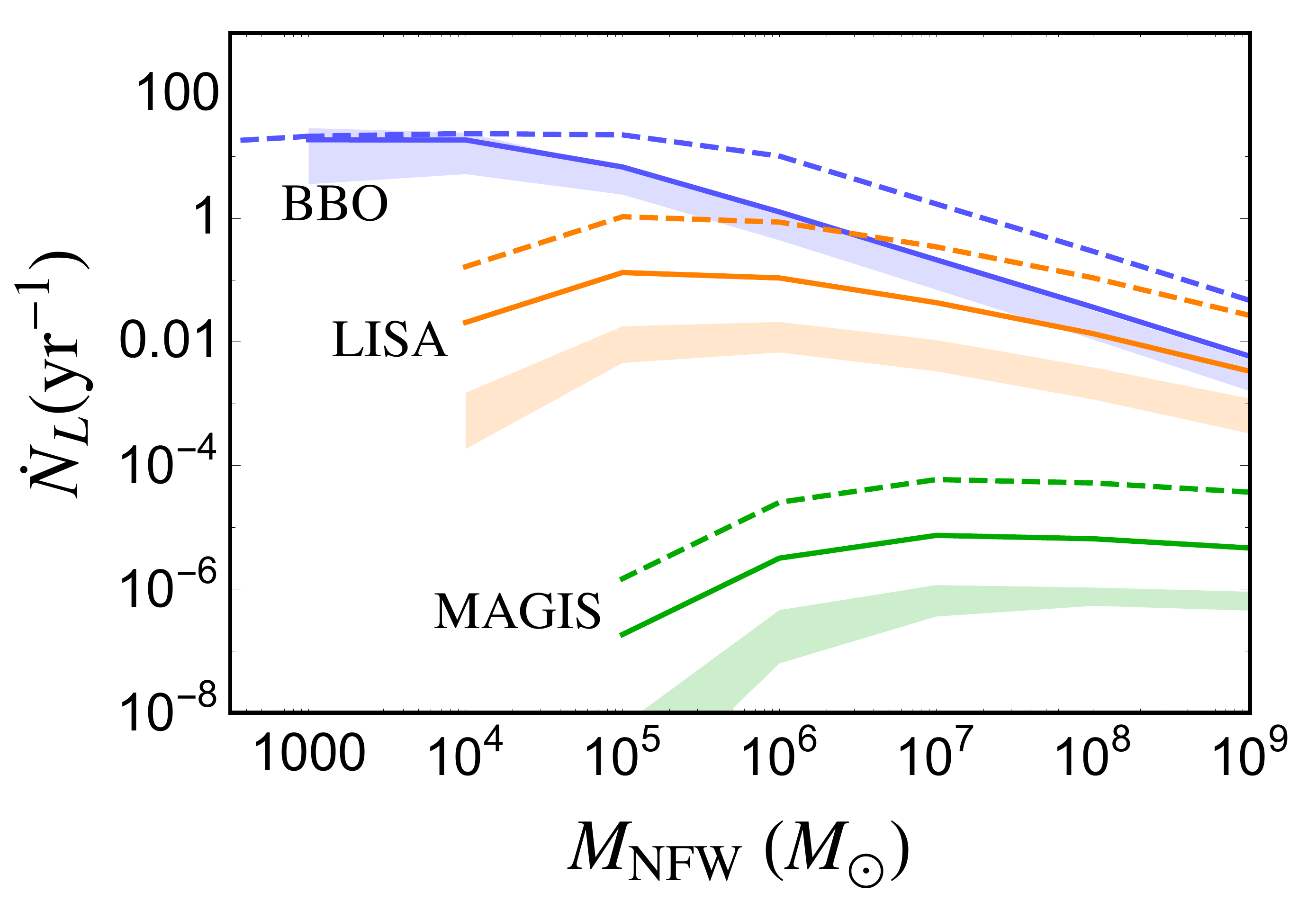}
\includegraphics[width=0.9\linewidth]{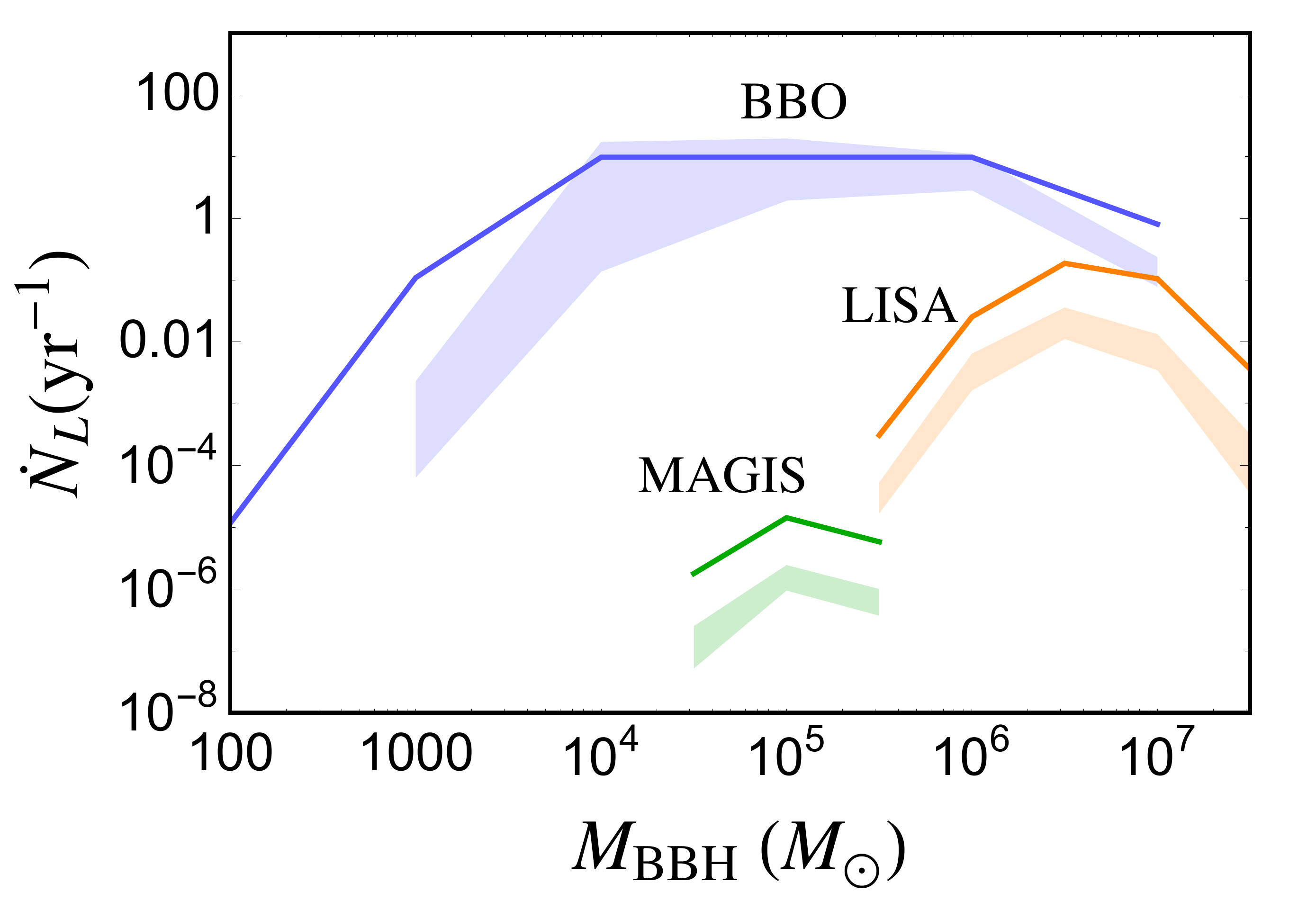}
\caption{ \label{fig:eventrate}
The number of lensing events per year $\dot{N}_L$ at BBO, LISA, and MAGIS, as functions of $\MNFW$ (top) or $\MBBH$ (bottom). The shaded bands are the range of optimistic and pessimistic $\dot{n}_s$, while the solid lines are from constant $\dot{n}_s$; their total event rates are normalized as in Table~\ref{tab:table1}. For comparison, dashed lines show results without a mass function; each $\MNFW$ comprises total $\Omega_{\rm DM}$. See more in text. Last one year of inspiral and $3\sigma$ log-likelihood lensing detection.
} 
\end{figure}

Table~\ref{tab:table1} shows total lensing events per year $\dot{N}_L$. Results are marginalized over $M_\NFW = 10^3 - 10^{10} \Msun$ with a mass function
\beq
\frac{ dn_l }{ dM_\NFW } \, \propto \, M_\NFW^{-2}
\label{eq:mnfwdist} \eeq
and summed for $\MBBH = 10^2- 10^8 \Msun$ with three models of $\dot{n}_s$. The power slope of a mass function is taken to be $-2$ for simplicity; heavier halos may contain abundant baryons that are not well described by NFW, while lighter halos' existence and properties are more model dependent. As for the three models of $\dot{n}_s$ (as a function of $\MBBH$ and $z_s$), two of them are taken from the models of massive black hole mergers in Ref.~\cite{Bonetti:2018tpf}; the most optimistic and pessimistic predictions are used.
Another model, as a simple reference, is constant $\dot{n}_s = 0.01 \ {\rm Gpc}^{-3} {\rm yr}^{-1}$ for all $\MBBH$ and $z_s$; this reference choice predicts similar total GW detection rates $\dot{N}_{\rm GW}$, as shown in the last three columns of Table.~\ref{tab:table1}. In all cases, BBH mergers are considered for $z_s \leq 10$ and $\MBBH = 10^2 \sim 10^8 \Msun$, where lighter BBHs have too small SNRs to contribute to $\dot{N}_L$ although they may contribute sizably to $\dot{N}_{\rm GW}$ (see \Fig{fig:eventrate} second panel).

Above all, in Table~\ref{tab:table1}, all three models of $\dot{n}_s$ predict that BBO can detect ${\cal O}(10)$ lensing events per year, while LISA barely detects a single event, and MAGIS and ET detect no event. Even though LISA and BBO have relatively large $\tau$, the number of relevant sources is not so large to start with (see the $\dot{N}_{\rm GW}$ column).

\begin{table}[t]
%\begin{ruledtabular}
\begin{tabular}{l cccccc}
\toprule
\textrm{Detector}&
\multicolumn{3}{c}{\textrm{$\dot{N}_L$}}&
\multicolumn{3}{c}{\textrm{$\dot{N}_{\text{GW}}$}}\\
 & const. & optim. & pessim. & const. & optim. & pessim. \\
%\colrule
\hline
BBO & 30 & 40 & 10 & 58 & 270 & 13   \\
LISA & 0.3 & 0.03 & 0.02 & 47 & 63 & 12  \\
MAGIS & \multicolumn{3}{c}{$< 10^{-5}$} & 25 & 187 & 9 \\
ET & \multicolumn{3}{c}{0} & 21 & 124 & 1 \\
\botrule
\end{tabular}
%\end{ruledtabular}
\caption{\label{tab:table1}
The expected numbers of lensing detections per year $\dot{N}_L$ and of total GW detections per year $\dot{N}_{\rm GW}$, at BBO, LISA, MAGIS, and ET. The results are marginalized over $\MNFW = 10^{3-10}\Msun$ with the mass function \Eq{eq:mnfwdist} and summed for $\MBBH = 10^{2-8} \Msun$ with three models of $\dot{n}_s$: constant $\dot{n}_s = 0.01 \, {\rm Gpc}^{-3}{\rm yr}^{-1}$, optimistic and pessimistic merger models of heavy BBHs~\cite{Bonetti:2018tpf}. Light BBH mergers are ignored.}
\end{table}

Which $\MNFW$ range has high event rates? In \Fig{fig:eventrate} upper panel, we show the event rates in log intervals of $\MNFW$ with the mass function. 
Most importantly, we conclude that the target range $\MNFW \lesssim 10^7 - 10^8 \, \Msun$ can be probed by diffractive lensing at BBO (and marginally at LISA). As discussed in \Sec{sec:MNFW} and \Fig{fig:Detectable2}, this range has the right scale radii $r_0$ that happens to coincide with the range of $r_F$ at these detectors. Although MAGIS and ET also have the right frequency scales, their SNRs are typically too small.
Notably, most BBO events are expected from light NFWs; ${\cal O}(10)$ events from light $\MNFW = 10^3 - 10^5 \Msun$,  ${\cal O}(1 - 10)$ from $\MNFW = 10^5 - 10^7 \Msun$, and smaller from heavier NFWs. LISA and MAGIS are relatively more sensitive to heavier NFWs, albeit with smaller event rates.

Figs.\ref{fig:tau} and \ref{fig:eventrate} also show an important feature of diffractive lensing: heavier NFWs yield smaller $\tau \propto \MNFW^{-0.8}$ (at low $z_s$). Therefore, unlike geometric-optics lensing, lighter NFWs are actually more sensitive. It is because the number density of heavier NFWs falls ($n_l \propto 1/\MNFW$) more quickly than the increase of the proper lensing cross section ($\sigma_l \propto \MNFW^{0.2}$). 
This is understood since the length scale of diffractive lensing is determined dominantly by $r_F$, not by $\MNFW$, since the $r_F$ range is much narrower than the $r_0$ range. For example, consider diffractive lensing by $\MNFW = 10^3\ \Msun$ and $10^9\ \Msun$ probed by a common $\MBBH=10^5\Msun$; even though their masses and $r_0$ differ sizably by $10^6$ and $\sim 300$ (\Eq{eq:rsMNFW}), the relevant range of $r_F$ is commonly fixed to be about $\sim 10$ (\Fig{fig:rFrange}) so that the lensing cross sections cannot differ by more than $\sim 10^2$.
This is why $\sigma_l$ is not so sensitive to $\MNFW$ that $\tau$ has a negative slope with $\MNFW$.\footnote{As an aside, if detection criterion is relaxed (say, $3\sigma$ to $2\sigma$), $\tau$ becomes steeper $\propto \MNFW^{-1}$, as the lighter NFW detection is more subject to the criterion.}

This is in stark contrast to usual geometric-optics lensing. For millilensing perturbations discussed in \Sec{sec:intro} and \App{app:MNFWmilli}, $n_l \sigma_l \propto \MNFW^{1.5 - 4}$ has a large positive slope with a mass so that light subhalos are inherently insensitive.  The strong lensing by a point mass $M$ is another example, where $r_E \gtrsim r_F$ makes $\sigma_l \propto r_E^2 \propto M$. But in this case, the power is canceled by that of $n_l \propto 1/M$ so that very light compact DM can also be probed with lensing, as mentioned. Diffractive lensing is more preferentially sensitive to lower masses.

Then, what does determine the lower range of $\MNFW$? \Fig{fig:tau} shows that, at low $z_s$ (only down to $\MNFW \gtrsim 10^2, 10^4, 10^6 \Msun$) can have sizable $\tau$ at BBO, LISA, and MAGIS.
As discussed in \Fig{fig:Detectable2}, it is either too long Fresnel length (for BBO and LISA with large enough SNRs) or too small SNR (for MAGIS and ET); light enough NFWs would have too small $r_0$ or too weak gravity to induce large enough diffractions. 
Moreover, the weaker gravity also limits the sensitivity at high $z_s$ for lighter NFWs.
The highest range of $z_s$ roughly scales with $\gamma(r_0)$, since SNR $\propto 1/z_s \gtrsim 1/\gamma$. For example, the ratio of $\gamma(r_0)$ between $\MNFW=10^5$ and $10^7 \Msun$ is about three (\Fig{fig:Detectable2}), and this roughly explains why $\MNFW=10^7\Msun$ can probe three times farther $z_s$, e.g. at LISA. Meanwhile, the decrease at small $z_s$ is  due to the small number of lenses and small $\kappa_0 \propto d_{\rm eff}$.

\Fig{fig:eventrate} (lower panel) also shows the event rates in terms of $\MBBH$ (with $\MNFW$ summed with the mass function). The largest $\tau$ is obtained for $\MBBH = 10^{6-7}\Msun$ at LISA, $10^{4-6} \Msun$ at BBO, and $\sim 10^5 \Msun$ at MAGIS. They are the mass ranges that typically produce largest SNRs. As expected, large SNR $\gtrsim {\cal O}(10^3)$ is needed to overcome small fractional changes of waveforms $\sim {\cal O}(
\gamma(r_0)) \sim {\cal O}(\kappa_0) \lesssim {\cal O}(10^{-3})$. Such a large SNR is readily obtained at LISA and BBO from heavy BBHs, while rarely at MAGIS, simply due to larger noise. Although the frequency range of ET is right to induce diffractive lensing by small NFWs (see \Fig{fig:Detectable2}), SNRs are just too small.

The variations between optimistic and pessimistic predictions are shown as shaded bands. They are only about $\sim 10$. But the predictions from the constant $\dot{n}_s$ (solid) at LISA and MAGIS tend to be larger (by about 10) even though they had similar $\dot{N}_{\rm GW}$. This tendency stems from that the massive-black-hole merger models predict more sources at higher $z_s$ so that LISA and MAGIS with smaller SNRs depend more sensitively on such distributions of source properties.

Lastly, the results without a mass function (dashed)
have almost the same shape as the solid lines but just a larger normalization by a factor $\sim 8$. One exception is at low $\MNFW$ range of BBO, where $\tau >1$ had to be cut off at $\tau = 1$ in our calculation.
These events are where multilensings of a single GW can occur. If the SNR is very large, even tiny lensing effects that might happen multiple times along the line of sight can all be counted. Such events may not be well detected as signals will be complicated, depending on many parameters of multilens environments. Using $\tau=1$ for such events means that we can always select out single-lensing events by, e.g., imposing stronger detection criteria for such events,  favoring the ones with single strong lensing and small perturbations.

This completes our study on the NFW DM subhalos to which weak diffractive lensing is applied.

%%%%%%%%%%%
\section{Generalization} \label{sec:discussion}

In this section, by working out lensing by power-law profiles, we not only demonstrate how readily one can estimate diffractive lensing in general (using our formalism), but also complete our discussions with strong diffractive lensing and the idea of measuring/distinguishing mass profiles.

%%%%%
\subsection{Lensing by power-law profiles} \label{sec:power-law}

Starting from a general power-law density profile 
\bea
\rho(x) &\=& \rho_0 x^{-k-1},  \qquad (0<k<2)
\eea
with $x=r/r_0$ for some scale $r_0$, we obtain 2D projected potentials 
\bea
\overline{\kappa}(x) &\=& \frac{2\kappa_0}{2-k} x^{-k}, \, \kappa(x) \= \kappa_0 x^{-k}, \, \gamma(x) \= \frac{k \kappa_0}{2-k}x^{-k},
\eea
with
\bea
\kappa_0 &\=& 4\pi d_{\rm eff} \rho_0 r_0 \mathrm{B} \left( \frac{1}{2}, \frac{k}{2} \right).
\eea
The range of $k$ makes the enclosed mass finite.

We fix the overall scale by specifying $\Mvir$. Further by choosing $r_0=r_E$, 2D projected potentials are simplified as
\bea
\overline{\kappa}(x) &\=& x^{-k}, \, \kappa(x) \= \frac{2-k}{2}x^{-k}, \, \gamma(x) \= \frac{k }{2}x^{-k},
\eea
now with $x = r/r_E$. The Einstein radius is fixed by $\Mvir$ as
\bea
    r_E &\=&  \left[ \frac{8\pi}{2-k} d_{\text{eff}} \,\rho_0 r_0^{k+1} \,\mathrm{B}\left(\frac{1}{2},\frac{k}{2}\right)\right]^{\frac{1}{k}}\, ,\\
    \rho_0 r_0^{1+k} &\=& \frac{200\rho_c(2-k)}{3}\left(\frac{3}{4\pi} \frac{M_\text{vir}}{200\rho_c}\right)^{\frac{1+k}{3}}\, ,
\eea
where $\rho_c = 3H_0^2/(8\pi)$ and $\mathrm{B}(x,y)= \Gamma(x)\Gamma(y)/\Gamma(x+y)$. Unlike NFW, $k<2$ profiles have non-negligible $r_E$ so that it is a useful length scale when it is comparable to the $r_F$ of GWs. 

\begin{figure}
\includegraphics[width=0.85\linewidth]{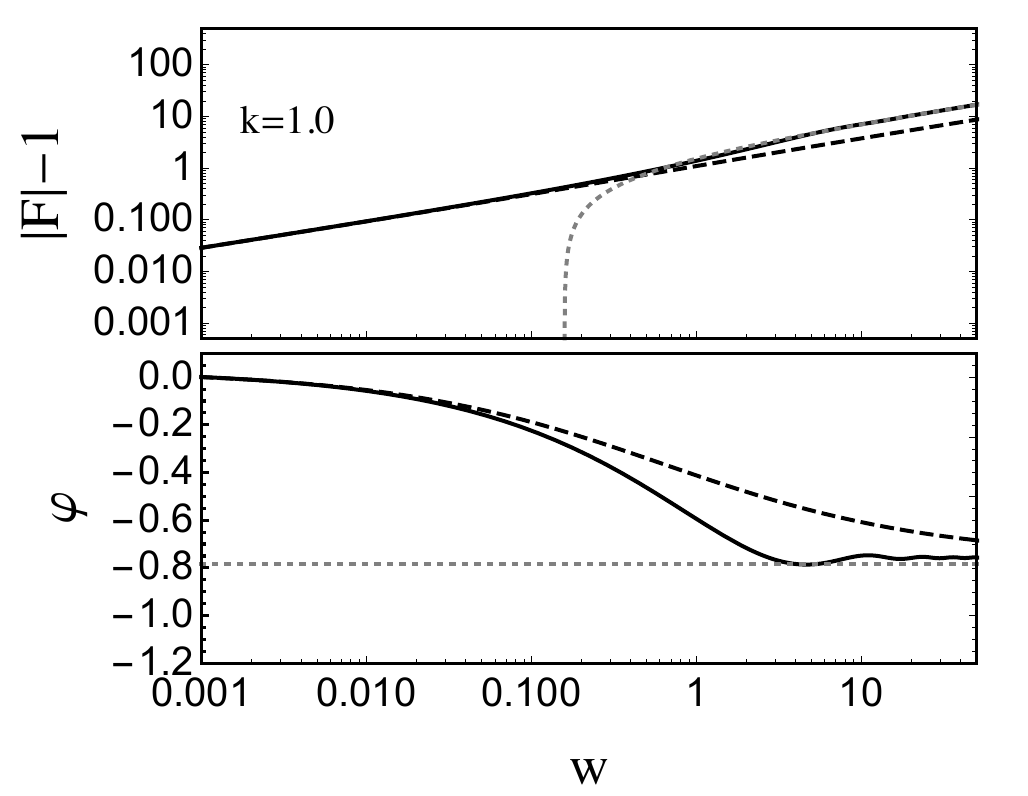}
\caption{\label{fig:Fpowerk1}
The amplitude and phase of $F(w)$ obtained by full calculation (solid), weak diffraction approximation \Eq{eq:powerF} (dashed), and strong diffraction approximation \Eq{eq:powerFstrong} (dotted) for a power-law profile with $k=1$. $x_s=0$ for simplicity.
} 
\end{figure}

For weak diffractive lensing which is valid for $w=2(r_E/r_F)^2\lesssim \text{min}(1,2/x_s^2)$ [\Eqs{eq:waveoptics2}{eq:generalcond}], our approximate results in terms of $\overline{\kappa}(x)$ and $\gamma(x)$ are
\begin{align} \label{eq:powerF}
F(w) &\,\simeq\, 1+ \frac{w}{i}\int_0^{\infty} d x \, x e^{i w \frac{x^2}{2}} \, x^{-k} \notag \\
 & \= 1 + 2^{-\frac{k}{2}}e^{-i \frac{k\pi}{4}}\Gamma\left(1-\frac{k}{2}\right) \,w^{\frac{k}{2}} \notag\\
 & \= 1+ 2^{-\frac{k}{2}} \Gamma\left(1-\frac{k}{2}\right) \, \overline{\kappa}\left(\frac{1}{\sqrt{w}}e^{i\frac{\pi}{4}}\right)\, ,
\end{align}
\begin{align} \label{eq:powerdF}
 \frac{dF(w)}{d\ln w} &\,\simeq\, \frac{w}{i}\int_0^{\infty} d x \, x e^{i w \frac{x^2}{2}} \, \frac{k}{2} x^{-k} \notag\\      
&\= 2^{-\frac{k}{2}}e^{-i \frac{k\pi}{4}}\Gamma\left(1-\frac{k}{2}\right) \, \frac{ kw^{\frac{k}{2}} }{2} \notag \\
        &\= 2^{-\frac{k}{2}}\Gamma\left(1-\frac{k}{2}\right) \gamma \left(\frac{1}{\sqrt{w}}e^{i\frac{\pi}{4}}\right)\, .
\end{align}
Here, integrals are evaluated exactly and the results agree with \Eqs{eq:meankFapprx}{eq:gammadFapprx} obtained from dominant supports.
Above the weak diffraction range, but still within $w<1/(2 x_s)$, strong diffractive lensing is described by \Eq{eq:strongdiff} which is calculated in this case as
\begin{align}\label{eq:powerFstrong}
    F(w) \,\simeq\, i^{-1/2} \sqrt{\frac{2\pi w}{1 - \kappa(1)-\gamma(1)}}  \= i^{-1/2} \sqrt{\frac{2\pi w}{k}}\, .
\end{align}

In \Fig{fig:Fpowerk1}, we compare $F(w)$ obtained by full calculation (solid), weak diffraction \Eq{eq:powerF} (dashed), and strong diffraction \Eq{eq:powerFstrong} (dotted) for $k=1$. Approximate results agree with the full results in their respective validity  ranges, confirming not only analytic calculations but also the validity ranges of weak/strong diffractions \Eqs{eq:waveoptics2}{eq:generalcond}. Weak diffraction starts to deviate at $w\gtrsim 0.1$ somewhat earlier than at $1$ since the Born approximation starts to break near $r_E$. Weak and strong diffractive lensing do have different slopes transitioning at around $w = 2 r_E^2/r_F^2 \simeq 1$ (the difference was explained in \Sec{sec:strong}), thus $r_E$ (existence and value) can be directly measured, effectively yielding $M_E = r_E^2/4d_{\rm eff}$ too. In the figure, $x_s=0$ for simplicity, but frequency independent results will arise for $w\gtrsim 1/2x_s$ with finite $x_s$, similarly to \Fig{fig:FdFex}. Although not shown, $\varphi(w)$ in this regime does not asymptote to zero (unlike the NFW case in \Fig{fig:FdFex}) because the relative times delays among multiple images remain there.

%%%%%
\subsection{Semianalytic estimation} \label{sec:SISestimation}
%\medskip
Using our analytic solutions, we estimate the detection prospects of diffractive lensing by power-law lenses. 

\begin{figure}[t]
\includegraphics[width=0.9\linewidth]{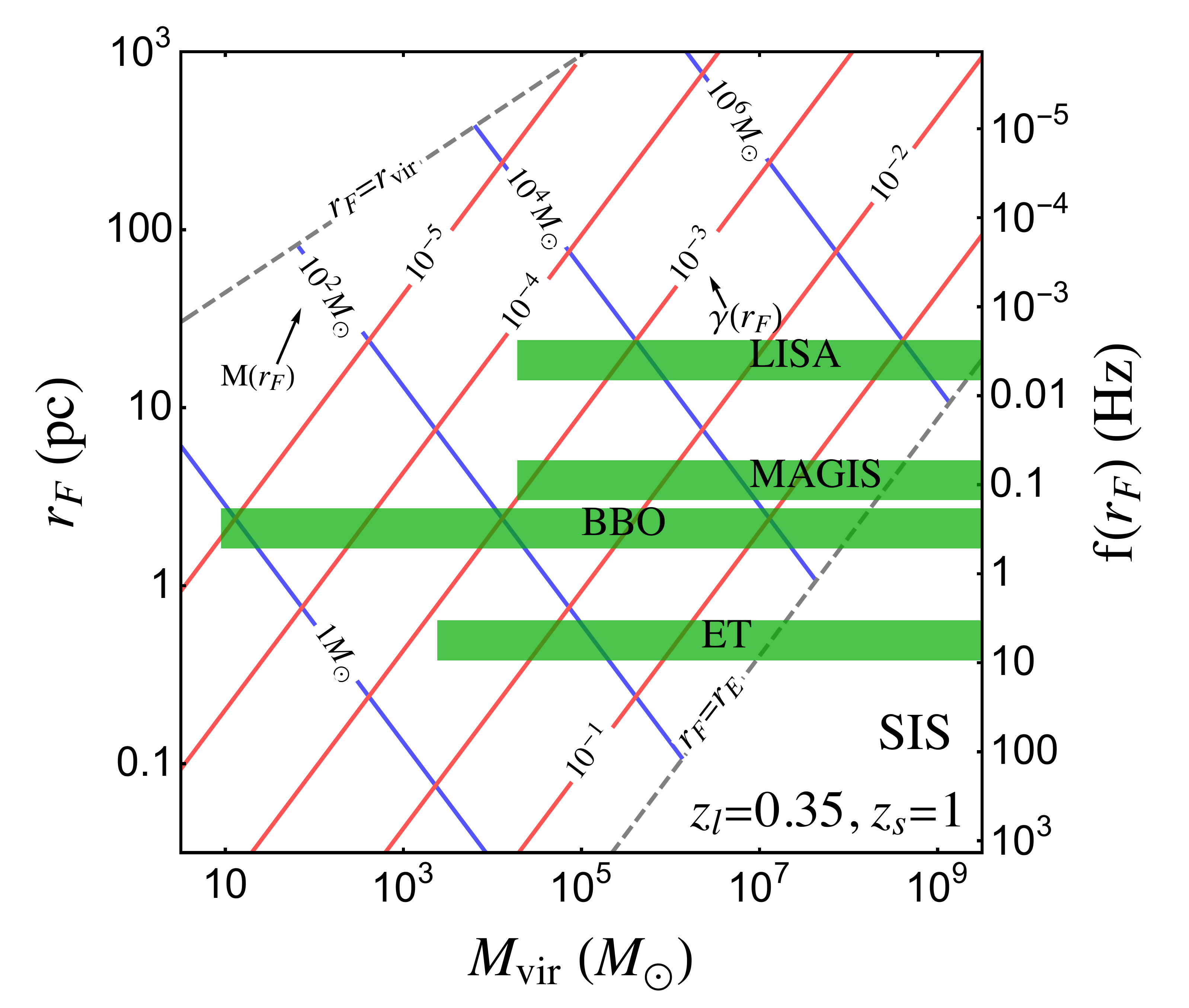}
\caption{ \label{fig:SISparam}
Same as \Fig{fig:Detectable2} but for SIS with $k=1$. Red contours show $\gamma(r_F)$, reflecting the required SNR for detection, and blue contours show the enclosed mass within $r_F$. Length scales, $r_{\rm vir}$ and $r_E$, are shown as dashed lines.} 
\end{figure}

To start off, as done for NFW, we estimate the relevant parameter space of the profile with $k=1$ in \Fig{fig:SISparam}. This is called the SIS profile, and is conventionally written in terms of the isothermal velocity dispersion $\sigma_v$ as $\rho(r)=\sigma_v^2/(2\pi r^2)$. 2D projected potentials are dimensionless 
\bea
\overline{\kappa}(x) &\=& 1/x, \quad
\kappa(x) \= 1/2x, \quad \\
\gamma(x) &\=& 1/2x \=  0.07 \left( \frac{\sigma_v}{1\, {\rm km/s}} \right)^2 \left( \frac{ d_{\rm eff} }{\rm Gpc} \right)\bigg(\frac{{\rm pc}}{r}\bigg),
\eea
with $x=r/r_E$, but scale parameters are rewritten as
\begin{align}
    & r_E = 4 \pi d_{\rm eff} \sigma_v^2  \= 0.14 \,{\rm pc} \times \left( \frac{\sigma_v}{1\, {\rm km/s}} \right)^2 \left( \frac{ d_{\rm eff} }{\rm Gpc} \right),
\end{align}
and the enclosed mass $M(r) \= \pi \sigma_v^2 r$
%\beq
%M(r) \= \pi \sigma_v^2 r \= 7.30 \times 10^{2} \Msun \left( \frac{\sigma_v}{1 \,{\rm km/s}} \right)^2 \left( \frac{r}{ \rm pc} \right)
%\eeq
within $r_E$ and $r_{\rm vir}$
\bea
M_E &=& 4 \pi^2 d_{\rm eff} \sigma_v^4 \= 1.02 \times 10^{6} \Msun  \left( \frac{\sigma_v}{10\, {\rm km/s}} \right)^4 \left( \frac{ d_{\rm eff} }{\rm Gpc} \right), \\
M_{\rm vir} &=& \frac{2}{\sqrt{50}} \frac{\sigma_v^3}{H_0} \= 9.39 \times 10^{8} \Msun \left( \frac{ \sigma_v}{10 \, {\rm km/s}} \right)^3.
\eea

\begin{figure}[t]
\includegraphics[width=0.75\linewidth]{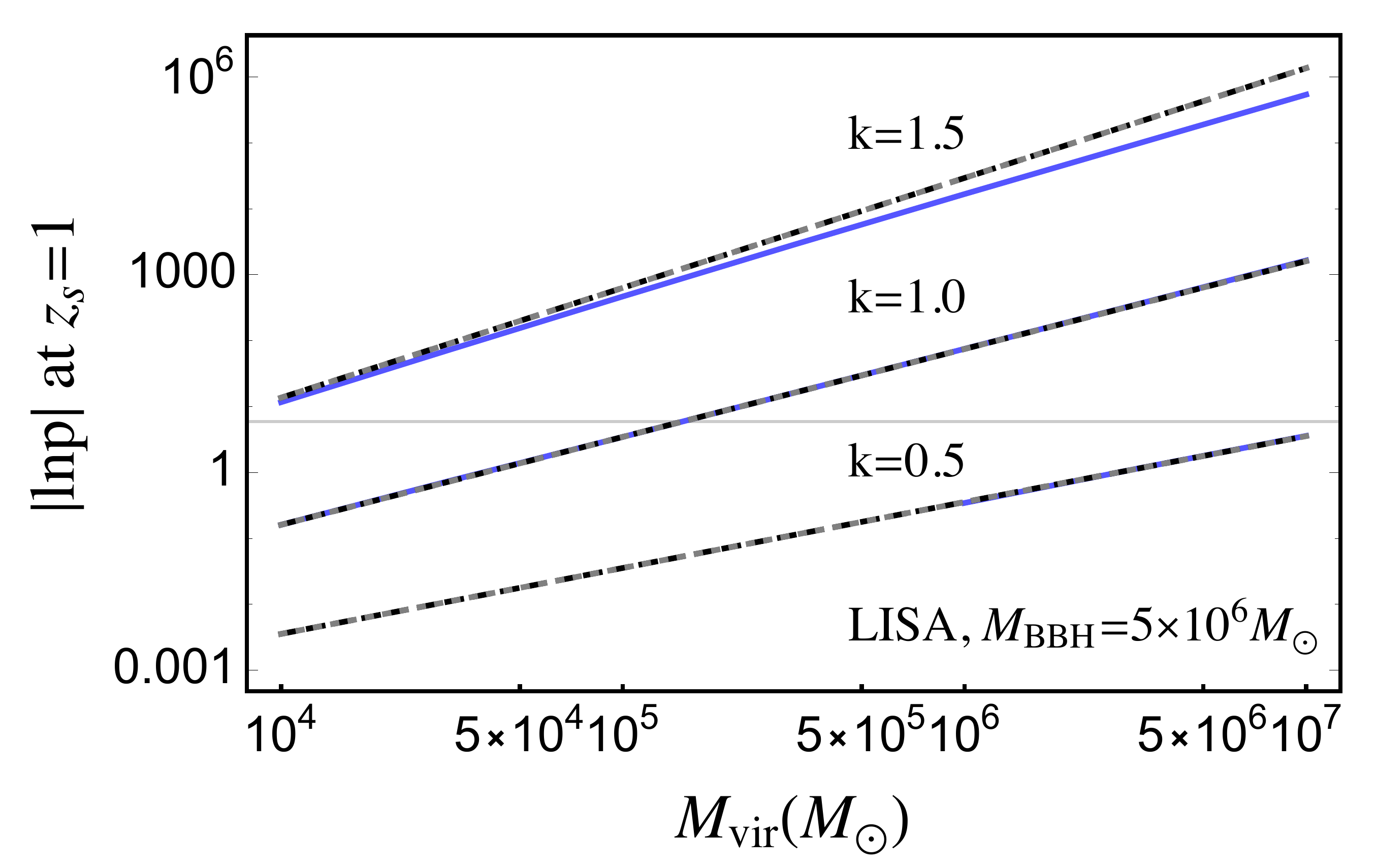}\\
\includegraphics[width=0.75\linewidth]{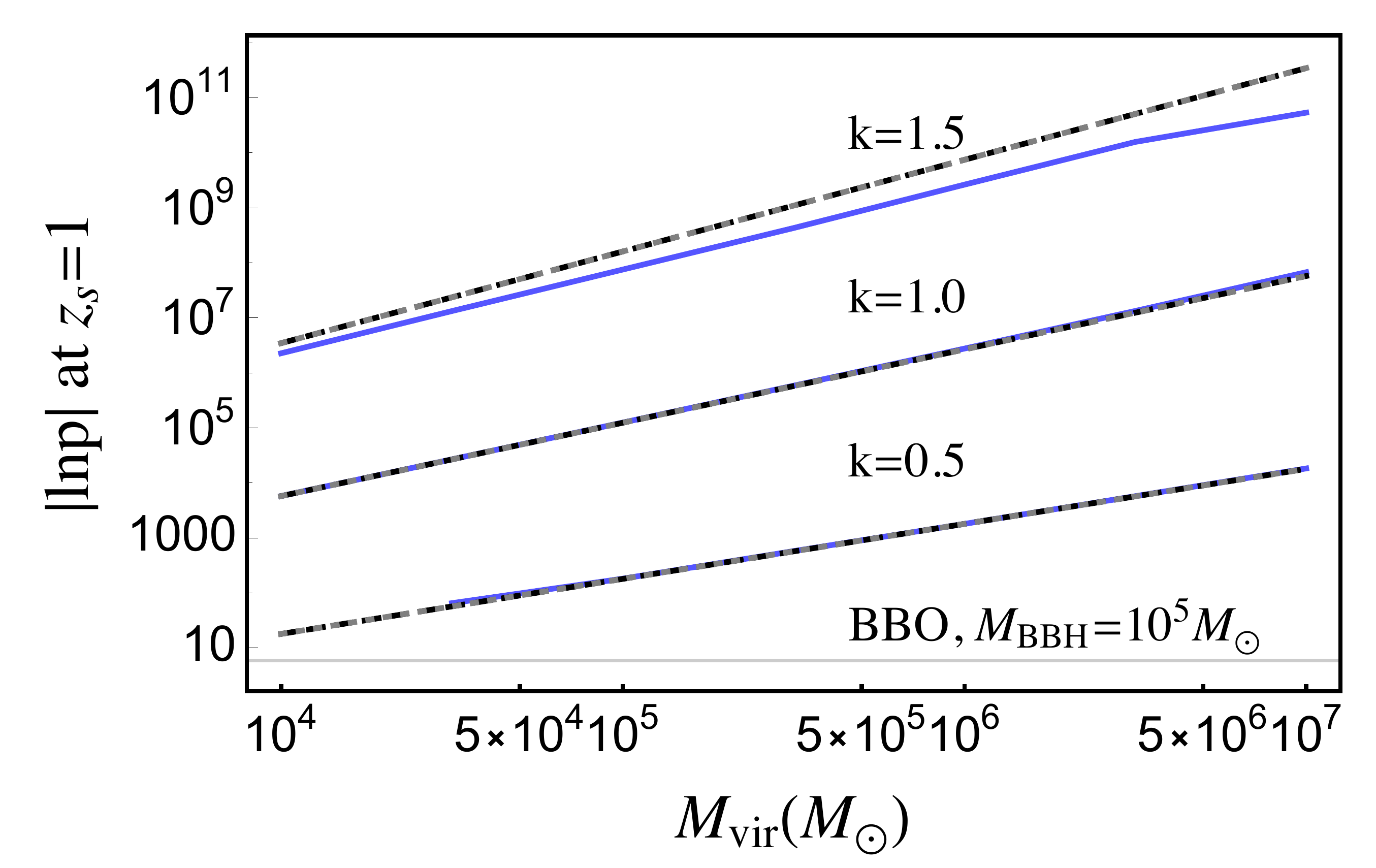}\\
\includegraphics[width=0.75\linewidth]{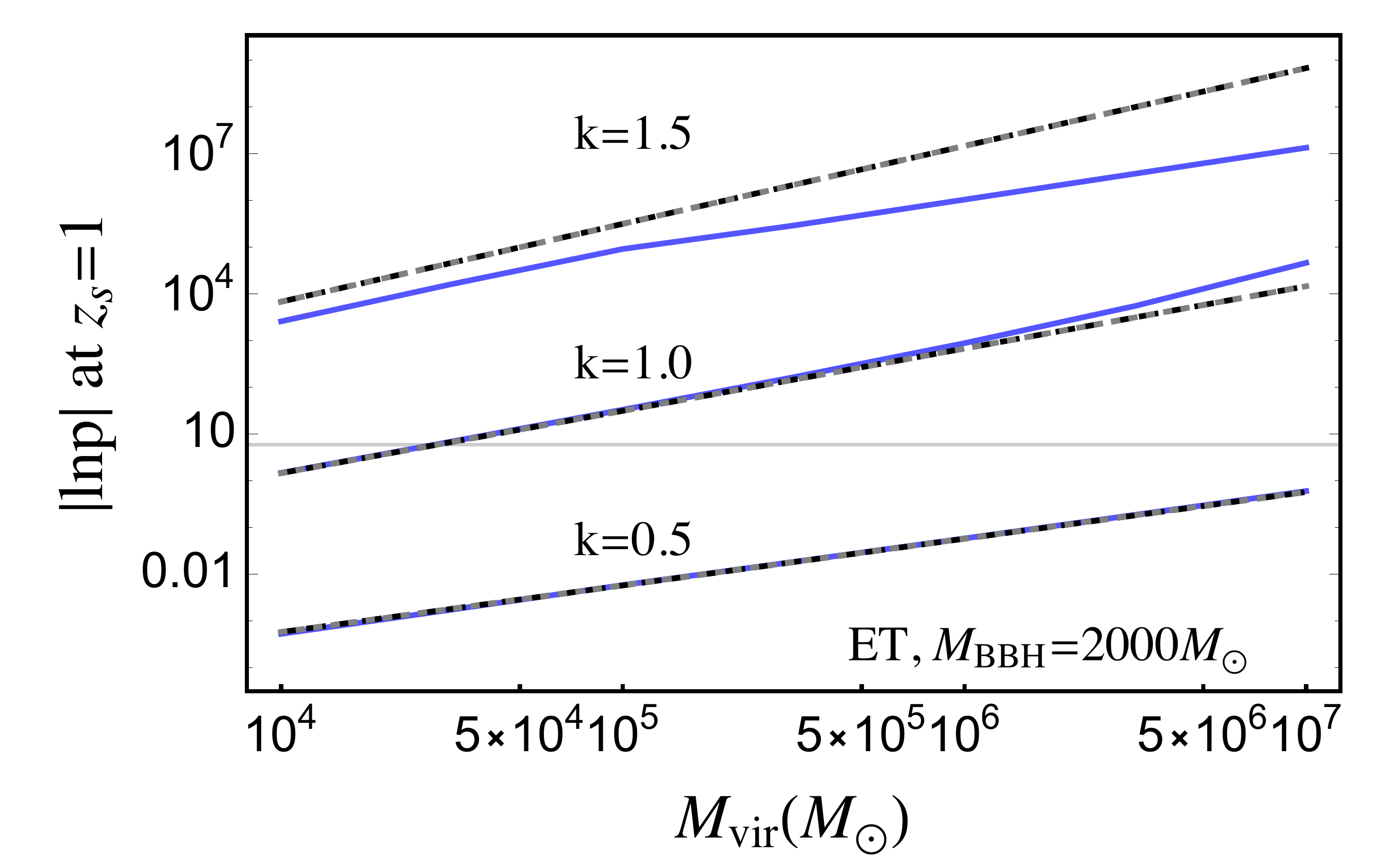}
\caption{ \label{fig:lnppower}
The comparison of $|\ln p|$ for detection obtained with full numerical (solid) and approximate weak diffraction \Eqs{eq:powerF}{eq:powerdF} (dashed). Also shown are estimations using only shear and SNR \Eq{eq:lnppowerrough} (dotted; which overlap with dashed). Each panel uses the BBH mass yielding maximum SNR. Horizontal lines denote the $3 \sigma$ threshold, $|\ln p |=5.914$. 
} 
\end{figure}

The detectable $M_{\rm vir}$ range is again estimated by the comparison of $\gamma(r_F)$ and SNR, with $r_F$ being the Fresnel length at the most sensitive frequency.
For example, ET ($r_F(f=10$ Hz)$ \,\sim\, 1 \,{\rm pc}$) with SNR $= {\cal O}(100)$ can probe a SIS lens as small as $M_{\rm vir} = 10^4 \Msun$ [or, $\sigma_v={\cal O}(1 {\rm km/s})$], corresponding to the enclosed mass $M(r_F) = 10 \Msun$ (blue solid). This estimation agrees with more dedicated calculations in Ref.~\cite{Dai:2018enj}, as the lower mass range is in the weak diffraction regime with $r_F \gg r_E\simeq 0.1 {\rm pc}$.

There are a few notable differences of \Fig{fig:SISparam} from NFW results of \Fig{fig:Detectable2}. The first is that  ET can probe smaller $\Mvir$ than MAGIS and LISA. This is because, for a given $M_{\rm vir}$, higher frequencies probe inner parts which now yield significantly larger shear, reflecting the steeper profile. Another is the relevance of the Einstein radius, which was essentially zero for NFW. This is further discussed in the following.

\begin{figure}[t]
\includegraphics[width=0.85\linewidth]{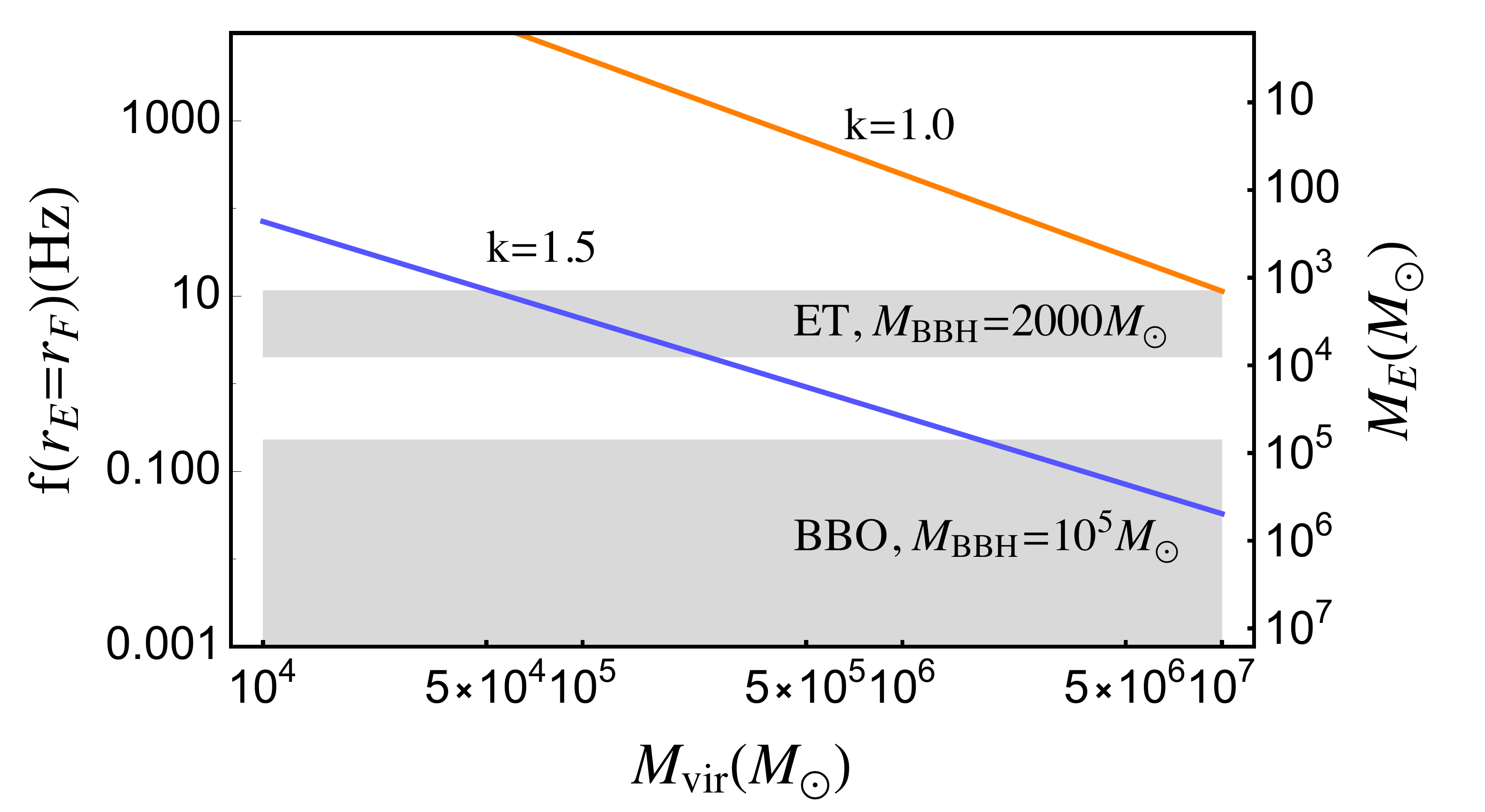}
\caption{ \label{fig:Mvirfreq}
The Einstein mass and the corresponding frequency for $r_F=r_E$, as a function of $\Mvir$. $k=1.0$ (orange) and $1.5$ (blue). The region above(below) each line is the strong(weak) diffraction regime. Shaded regions represent the chirping frequency ranges measured at given detectors. $z_s=1, z_l=0.35$. 
} 
\end{figure}

Further, we can estimate somewhat more accurately, but still much more easily than full numerical analysis. Using weak diffraction results ---\Eqs{eq:powerF}{eq:powerdF}---, we calculate $\ln p$ for detection by minimizing with respect to $A$ and $\phi_c$. This result is compared with full numerical result in \Fig{fig:lnppower}. They agree well for most $\Mvir$ and $k$, but deviates in the heavy mass region of large $k$ are due to strong diffractive lensing. As shown in \Fig{fig:Mvirfreq}, for given $\Mvir$, the larger the $k$, the larger the $M_E$ so that strong diffraction becomes more relevant from lower frequencies. In this region, the frequency slope $w^{1/2}$ \Eq{eq:powerFstrong} is steeper (shallower) than that of weak diffraction $w^{k/2}$ \Eq{eq:powerF} for $k<1 \,(k>1)$\footnote{
The turnover can be more accurately found to be $k\simeq 1.3$ using \Eqs{eq:powerF}{eq:powerFstrong}.}
 so that full results are stronger (weaker). In addition to these results, dotted lines show much simpler estimations based solely on shear and SNR [motivated in \Sec{sec:MNFW} and supported rigorously in \Eq{eq:lnpapprx}]
\beq \label{eq:lnppowerrough}
|\ln p| \,\simeq\, \alpha \left({\rm SNR}\times\gamma(r_F(f_*))\right)^2,
\eeq
where $\alpha = {\cal O }(0.1)$ reproduces the analytic results. In all, \Fig{fig:lnppower} confirms our analytic results and demonstrates how readily one can estimate diffractive lensing using our formalism.

%  

%%%%%
\subsection{Peeling off profiles}  \label{sec:peelprofiles}

\begin{figure}
\includegraphics[width=0.75\linewidth]{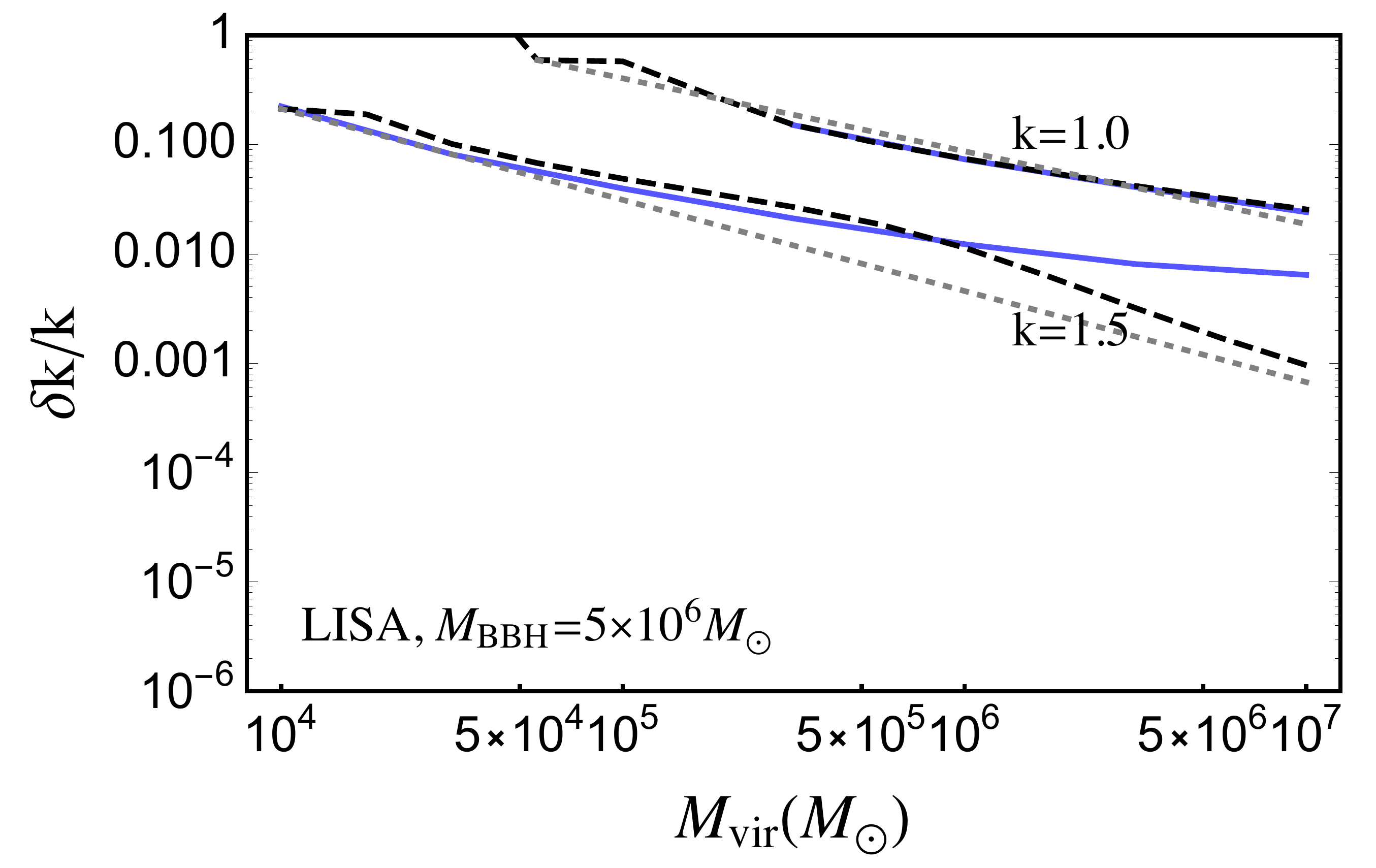}\\
\includegraphics[width=0.75\linewidth]{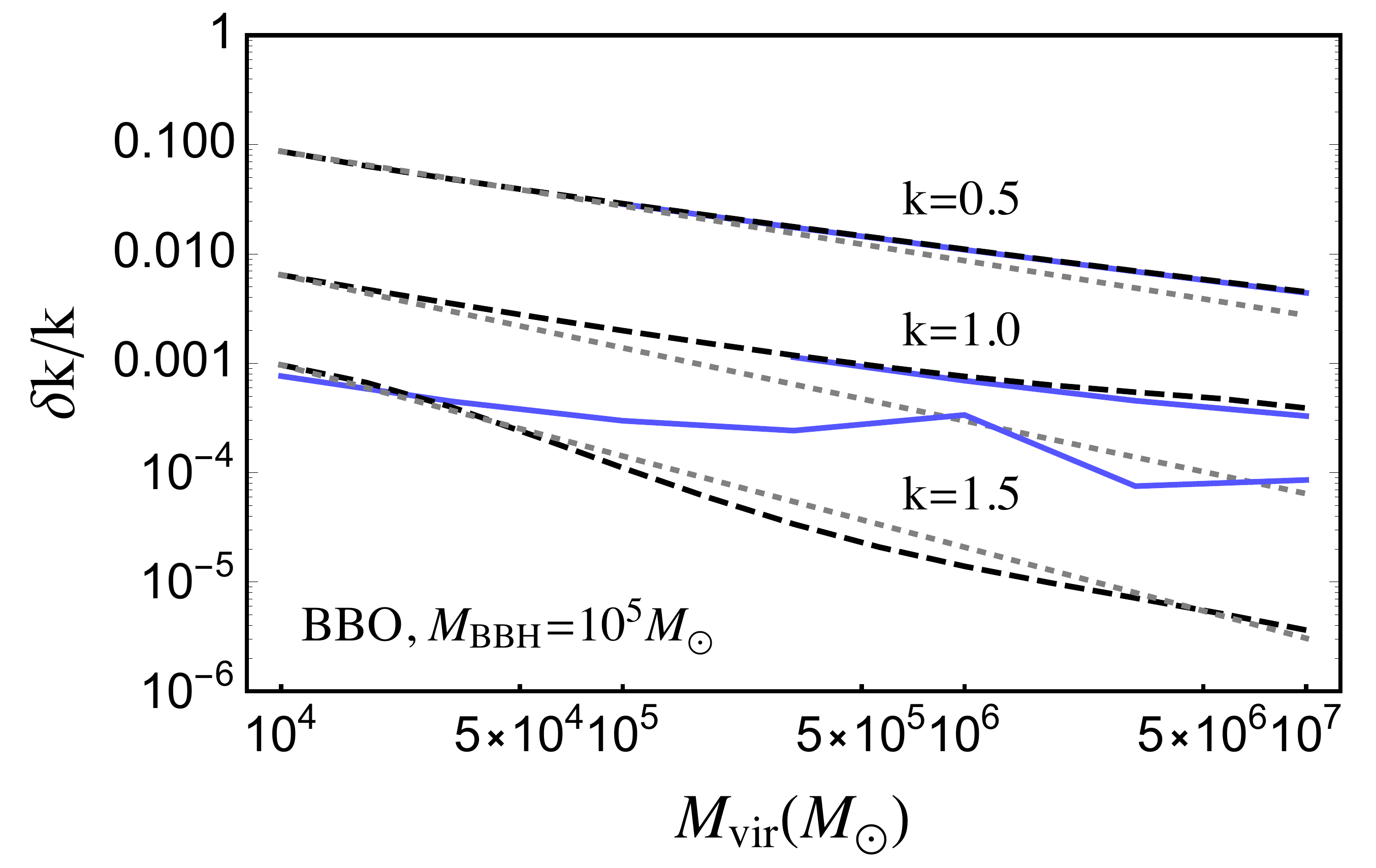}\\
\includegraphics[width=0.75\linewidth]{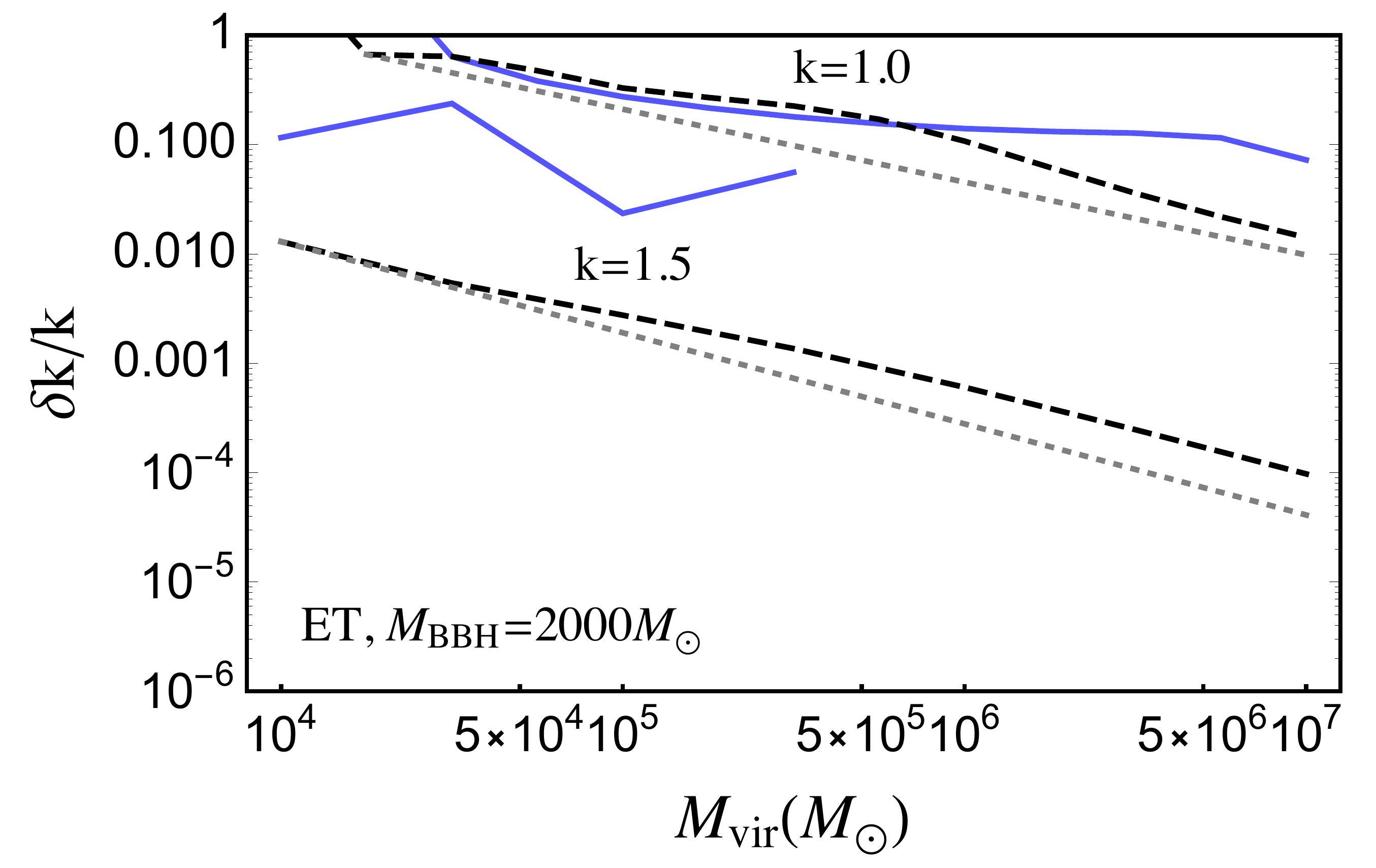}
\caption{ \label{fig:kmeasurement}
Same as \Fig{fig:lnppower} but for profile measurement accuracies represented by $\delta k/k$. Cases with $\delta k/k>1$ are not shown.
} 
\end{figure}

It was advocated that our formalism in terms of 2D potentials makes it clear what it means to measure the mass profile with a single diffractive lensing event. The basic idea is simple: different profile slopes $k$ result in different frequency dependence during the probe of a successively smaller length scale. 
As a simple demonstration of this exciting possibility, we estimate the measurement accuracy of the slope $k$.

Similarly to detection estimates, we calculate $\ln p$, but this time including $k$ as a fitting parameter (in addition to $A$ and $\phi_c$). We define the measurement accuracy $\delta k$ as the variation of $k$ with respect to the true $k_0$ that yields $|\ln p| = 5.914$. In \Fig{fig:kmeasurement}, we show the results, again obtained from full numerical, weak diffraction analytic, and shear-times-SNR. Above all, different calculations agree well if weak diffraction dominates (for small $k$ and small $\Mvir$). Measurement accuracies are good as long as lensing can be detected. Basically,  in the weak diffraction regime, the heavier or the steeper the lens is, the more accurate measurement or distinction of profiles.

Notably, full numerical results deviate more significantly and yield much worse results in \Fig{fig:kmeasurement}, compared to the detection prospects in \Fig{fig:lnppower}. This is an important effect of strong diffractive lensing, qualitatively different from weak diffraction. Strong diffraction has universal frequency dependence $w^{1/2}$ \Eq{eq:powerFstrong} independent of the power $k$ (as discussed carefully in \Sec{sec:strong}) was due to the breaking of the scale invariance by an Einstein ring. As a result, different profiles are harder to distinguish; detection itself was more  robust because it is essentially the comparison of power $k$ and flat potentials. Thus, peeling off profiles is possible only with weak diffractive lensing.

Since our analysis on diffractive lensing ignores the parameter degeneracies between the lens profile and GW waveform, there might be some overestimation in the profile measurement accuracy. But, in practice, the nonzero impact parameter $x_s$ which induces the diffraction-to-geometric-optics transition might be able to resolve some degeneracy between the profile and the GW waveform parameters as discussed in \Sec{sec:criteria}. 

%%%%%
\subsection{Core-vs-cusp}

If GW diffractive lensing can probe mass profiles, can it resolve the core-vs-cusp problem? There exist observational evidences that DM halos may contain flat cores of ${\cal O}(0.1-1)$ kpc radius~\cite{Bullock:2017xww,Moore:1994yx} rather than cuspy NFW $\propto 1/r$. Such cores would change lensing effects at the corresponding frequencies. But this length scale is too large, corresponding to too low frequencies $f \lesssim 10^{-4}$ Hz (\Fig{fig:rFrange}) for chirping GWs to be relevant; the LISA's most sensitive frequency range was $\sim 0.003$ Hz. It is currently more of a problem of halos rather than of subhalos. Whether this problem persists to smaller length scales (smaller DM-dominated halos) is not certain, and it is this question that can be answered by observations of GW diffractive lensing.

%%%%%%%%%%%
\section{Summary} \label{sec:summary}

First, we have developed a formalism for weak and strong diffractive lensing and solved it analytically. As a result, the complex lensing integral is evaluated in terms of much simpler 2D-projected potentials. In particular, the frequency dependence of weak lensing turns out to be due to shear of a lens at the Fresnel length $r_F \propto f^{-1/2}$. These results make not only  underlying physics of diffraction clearer but also its estimation much easier, as discussed and demonstrated throughout this paper. Moreover, the idea of measuring mass profiles became concrete. 

We have also derived the condition or the validity range of diffractive lensing. It turns out that there exist two different phases of diffraction: weak and strong. They are separated by the Einstein radius, outside of which is approximately scale invariant leading to $|F(w)|-1 \propto w^{k/2}$ (for power-law profiles) inside of which has only azimuthal symmetry leading to universal $|F(w)| \propto w^{1/2}$. The innermost range of diffractive lensing is determined by properties of a caustic (multi-imaged cases) or by the source location. 

Applying these, we have shown that NFW subhalos of $\MNFW \lesssim 10^7 \Msun$, which cannot be probed with existing methods,  can be detected individually with GW diffractive lensing. Detection prospects are ${\cal O}(10)$ events per year at BBO and less at LISA, limited mainly by small merger rates and large required SNR $\gtrsim 1/\gamma(r_0) \sim 10^3$. This mass scale is sensitive because the corresponding scale radius $r_0$ happens to be comparable to the range of $r_F$ at future GW detectors. Notably, unlike strong lensing observables, the scale of diffractive lensing is dominantly fixed by $r_F$ rather than $r_0$ (or the lens mass) so that it can be relatively more sensitive to lighter lenses.

Further, we have applied our formalism to readily estimate the detection and profile measurements for general power-law potentials. This application also clarifies the features of strong diffractive lensing and makes the idea of measuring mass profiles concrete. Just as the shear field measured from galaxy shape distortions is used to measure galactic profiles and matter power spectrum, GW diffractive lensing can potentially be used to measure small-scale shear and reveal the particle nature of DM roaming in the subgalactic scale.

%%%%%%%
%\section*{Acknowledgements}
%\begin{acknowledgments}
\bigskip
{\bf Acknowledgments.}
We thank Liang Dai, TaeHun Kim, Chan Park, and Arman Shafieloo for useful conversations.  We are also grateful to anonymous referee for a thorough review and constructive comments. We are supported by Grant Korea NRF-2019R1C1C1010050, 2019R1A6A1A10073437, 
and SJ also by POSCO Science Fellowship. We also appreciate APCTP for its hospitality during completion of this work.
%\end{acknowledgments}

%%%%%%%%%%%%%%%
\appendix

%%%%%
\section*{Appendix}

%%%%%
\subsection{$\MNFW$ scaling of millilensing perturbation} \label{app:MNFWmilli}

We estimate the sensitivity of millilensing perturbation observations on $\MNFW$. In particular, we aim to derive the dependence on the mass and the lower mass range, both of which can be contrasted with those of diffractive lensing.

The flux ratio anomaly is the most sensitive observable of millilensing perturbation; it is a second-derivative of the $\hat{T}_d$ surface~\cite{Mao:1997ek}. The typical requirement of $\gtrsim 10\%$ flux perturbation $\Delta \mu / \mu$ by NFW subhalos~\cite{Metcalf:2001es,Metcalf:2004eh,Xu:2011ru} is translated to the requirement of subhalo's $\kappa(x)$ as
\beq
\frac{\Delta \mu}{\mu} \, \simeq \, \kappa(x) \, \gtrsim \, 0.1,
\eeq
leading to maximum possible $x$ (using \Eqs{eq:psiNFW}{eq:curFNFW})
\beq
x \, \lesssim \, x_{\rm max} \, \simeq \, 2 \exp\left( - \frac{0.1}{3\kappa_0(\MNFW)} - \frac{1}{2} \right).
\eeq
Using $\kappa_0 \propto \MNFW^{0.18}$ \Eq{eq:kappa0} and $r_0^2 \propto \MNFW^{0.82}$ \Eq{eq:rsMNFW}, the lensing cross section $\sigma_l = \pi r_0^2 x_{\rm max}^2$ scales with the mass as
\beq
\frac{ d \ln \sigma_l}{d \ln \MNFW} \, \simeq \, 0.82 \+ 0.18 \left( \frac{2}{3} \frac{0.1}{\kappa_0(\MNFW)} \right) \, \simeq \, 5 - 2.5
\eeq
for $\MNFW = 10^7 - 10^9 \Msun$ (having $\kappa_0(\MNFW) = 0.003 - 0.008$), respectively. Thus, $n_l \sigma_l \propto \MNFW^{4 - 1.5}$ scales rapidly with the mass. Although heavier masses are subject to larger shot noise, this scaling inherently limits the sensitivity to light NFWs. If the profile were more compact as for SIS or pseudo-Jaffe with a power-law $\kappa(x) \propto 1/x$, the mass dependence would have been shallower as $n_l \sigma_l \propto M_{\rm vir}^{1/3}$. As emphasized, this positive scaling slope is in stark contrast with the negative slope of diffractive lensing (which makes GW diffraction more suitable to probe light NFWs).

Now, how small $\MNFW$ can be detectable with sizable probabilities? The average 2D-projected separation of NFW subhalos within the Einstein radius 5 kpc of a galaxy is about ${\cal O}(0.1) \, r_0$ (if a whole DM abundance is in the form of subhalos and is uniformly distributed).
So, by requiring $x_{\rm max} \gtrsim 10^{-3} - 10^{-2}$ for sizable optical depths, we obtain $\MNFW \gtrsim 10^7 - 10^9 \Msun$. This is the current lower limit \cite{Hezaveh:2016ltk,Asadi:2017,Nierenberg:2014cga,Fadely:2012}, which will not be improved significantly in the future.

%%%%%%%
\subsection{Range of diffractive lensing near a caustic} \label{app:caustic}

Near a caustic, time delays between the images formed just around corresponding critical lines are very small. Thus, very high frequency is  needed to reach the geometric-optics regime. We quantify this condition.

Start from a dimensionless time delay in \Eq{eq:hatTd} ($x= r/r_E$)
\beq
\Thatd({\bm x}, {\bm x}_s) \= \frac{1}{2} \left| {\bm x}-{\bm x}_s \right|^2 \- \psi(x),
\eeq
which appears in the path integral as $\int d^2 {\bm x} \, \exp[ i w \Thatd({\bm x}, {\bm x}_s)]$. The locations of geometric-optics images are stationary points, yielding the lens equation
\beq
\Thatd^\prime \= 0 \quad \leftrightarrow \quad  {\bm x}_s \= {\bm x} - \psi^\prime(x).   
\eeq
For given ${\bm x}_s$ with $x_s >0$, images can form in either side. Removing the vector notation and using $x>0$, we obtain two lens equations
\beq
x_s \= x - \psi^\prime (x), \qquad x_s \= -x + \psi^\prime (x).
\eeq
At the caustic $x_s=0$, images are formed at the critical line $x_t$ (in this case, the Einstein radius $x_t=x_E=1$)
\beq
x_t \= \psi^\prime(x_t),
\eeq
and the two solutions are connected to form an Einstein ring. Near a caustic with $x_s \ne 0$, two image locations are $x_t + \delta x$ and $-x_t + \delta x$ satisfying
\beq
x_s \= \delta x - \psi^{\prime \prime}(x_t) \delta x \qquad \leftrightarrow \qquad \delta x \= \frac{x_s}{1- \psi^{\prime \prime}(x_t)}.
\label{eq:deltax} \eeq
Thus, one image (in the same direction) is slightly outside the critical line, while the other (in the opposite direction) is slightly inside. Note that $\delta x$ and $x_s$ are proportional to each other.

The dimensionless time delay of each image is
\bea
\Thatd(x_t &+& \delta x) \,\simeq\, \Thatd(x_t) \+ \Thatd^{\prime}(x_t) \delta x + \cdots \\
&=& \frac{1}{2} ( x_t - x_s )^2 - \psi(x_t) + \big( (x_t - x_s) - \psi^\prime(x_t)  \big) \delta x,  \notag
\eea
and
\bea
\Thatd(-x_t &+& \delta x) \,\simeq\, \Thatd(-x_t) \+ \Thatd^{\prime}(-x_t) \delta x + \cdots \\
&=& \frac{1}{2} ( x_t + x_s )^2 - \psi(x_t) + \big( (x_t + x_s) - \psi^\prime(x_t)  \big) \delta x. \notag
\eea
The relative time delay is then
\bea
\Delta \Thatd &\=&  \Thatd(- x_t + \delta x) - \Thatd(x_t + \delta x) \\
&\,\simeq\,& 2 x_t x_s + 2 x_s \delta x \= 2 x_t x_s + {\cal O}(x_s^2).
\eea
Thus, diffraction occurs inside the Einstein radius if 
\beq
w \,\lesssim\, \frac{1}{2 x_t x_s} \qquad \leftrightarrow \qquad r_F \,\gtrsim \, 2 \sqrt{r_E r_s}
\eeq
rather than $w \lesssim 2/x_s^2$ (or $r_F \gtrsim r_s$) outside the Einstein radius. 
This is the innermost range of (strong) diffractive lensing discussed in \Eq{eq:generalcond}.

%%%%%
\subsection{Single image of diffractive lensing} \label{app:singleNFW}

We prove that diffractive lensing is single-imaged and that the image is always magnified as shown in \Fig{fig:FdFex}. The proofs are based on existing theorems and logics for general lensing properties (see e.g.,~\cite{Schneider:1992}). 

Each image is associated with an index characterizing whether it is located at an extremum or a saddle point of $\hat{T}_d$ surface. Define the angle $\varphi$ of the gravity force field on the lens plane as $\nabla \hat{T}_d \propto (\cos \varphi, \sin \varphi)$. The index can be defined as the loop integral of $\varphi$ around the image: $\frac{1}{2\pi} \oint_C d\varphi$ = $+1$ for a maximum or a minimum and $-1$ for a saddle. Index theorem says that a closed integral along an arbitrary loop is the sum of all enclosed indices
\beq
\frac{1}{2\pi} \oint_C d\varphi \= n_{\rm max} + n_{\rm min} - n_{\rm saddle},
\eeq
where the total number of images is $n = n_{\rm max} + n_{\rm min} + n_{\rm saddle}$.
Since $\hat{T}_d$ has an absolute minimum (corresponding to the minimum travel time), $n_{\rm min} \geq 1$. 

In the diffractive regime sufficiently far away from a lens, (1) $A \to I$ identity, and (2) $\nabla \hat{T}_d$ is radial. The latter implies $\frac{1}{2\pi} \oint_C d\varphi = 1$. The former implies $\tr A >0$ and $\det A >0$ so that all images are of the minimum-type (a saddle-type has $\det A <0$ and a maximum $\tr A <0$). Therefore, $n = n_{\rm min} =1$; diffractive lensing produces a single image, of the minimum-type. 

The (1) also implies $\tr A = 2(1-\kappa) >0$ (with $\kappa>0$) and $\det A = (1-\kappa)^2 -\gamma^2 >0$ so that $\gamma < 1- \kappa \leq 1$, hence $\det A <1$. Thus, the magnification of the single image is $\mu = 1/ \det A >1$, always magnified.

%%%%%%%%
\subsection{Formulation of $\ln p$} \label{app:formula-lnp}

The inner product between two time domain waveforms, $h_1(t)$ and $h_2(t)$, is defined as
\begin{equation}
    (h_1|h_2) = 4\ \text{Re} \int_0^{\infty} df \frac{\tilde{h}_1^*(f)\tilde{h}_2(f)}{S_n(f)}\ ,
\end{equation}
where $\tilde{h}_1(f)$, $\tilde{h}_2(f)$ are the Fourier transform of the time domain waveform and $S_n(f)$ is the noise spectral density of the detector.
For a detector output $s(t)$ and a waveform template $h_{\lambda_1,\lambda_2,\cdots}$, where $\lambda_1,\lambda_2,\cdots$ are free parameters of the template, the best-fit waveform $h_\text{BF}$ is the waveform that minimizes the inner product
\begin{equation}
    \mathcal{D}=(s-h_{\lambda_1,\lambda_2,\cdots}|s-h_{\lambda_1,\lambda_2,\cdots}).
\end{equation}
The lensed gravitational waveform in frequency domain $\tilde{h}_L$ is given by
\begin{equation}
    \tilde{h}_L(f) = F(f) \tilde{h}(f),
\end{equation}
where $\tilde{h}(f)$ is an ordinary gravitational wave without lensing effects generated by compact binary coalescence. 

Suppose the signal $s(t)$ is well described by the lensed waveform $\tilde{h}_L(f)=F(f)\tilde{h}_{\lambda_1^0,\lambda_2^0,\cdots}(f)$ and we have unlensed template $\tilde{h}_{\lambda_1,\lambda_2,\cdots}(f)$.
Then the best-fit waveform $h_\text{BF}$ is given by minimizing
\begin{eqnarray}
    \mathcal{D}=&&(h_L-h_{\lambda_1,\lambda_2,\cdots}|h_L-h_{\lambda_1,\lambda_2,\cdots}) \nonumber \\
    =&&(Fh_{\lambda_1^0,\lambda_2^0,\cdots}|Fh_{\lambda_1^0,\lambda_2^0,\cdots})- 2(Fh_{\lambda_1^0,\lambda_2^0,\cdots}|h_{\lambda_1,\lambda_2,\cdots})\nonumber\\
    &&+(h_{\lambda_1,\lambda_2,\cdots}|h_{\lambda_1,\lambda_2,\cdots})
\end{eqnarray}
in the parameter space $\lambda_1,\lambda_2,\cdots$.
In general, the parameter space includes total mass, mass ratio of the compact binary, luminosity distance to the source, and etc. 
However, for simplicity of analysis, we consider only three parameters; constant phase $\phi_c$, overall amplitude $A$, and coalescence time $t_c$.  
Then, the lensed waveform and the template waveform can be written as
\begin{eqnarray}
    && \tilde{h}_L(f)=F(f)\tilde{h}_0(f) \\
    && \tilde{h}(f)=\tilde{h}_0(f)A e^{i\phi_c+2\pi i f t_c }\ ,
\end{eqnarray}
where the waveform $\tilde{h}_0(f)$ contains all the other parameter dependence of GW waveform.
Here, we set $\phi_c^0=t_c^0=0$ since they can be arbitrary.
Then, $\mathcal{D}$ is given by
\begin{equation}\label{eq:hDistance}
    \mathcal{D}=(F h_0|Fh_0 )-2 A (F h_0|h_0 e^{i  \phi_c+2\pi i f  t_c })+A^2 (h_0|h_0)\, .
\end{equation}
We can solve the minimization problem of Eq. (\ref{eq:hDistance}) analytically for $\phi_c$ and $A$.
If $\mathcal{D}$ is minimized for $\phi_c$ and $A$, it satisfies
\begin{eqnarray}
    \frac{\partial \mathcal{D}}{\partial \phi_c}&&  = -i A \left[e^{i \phi_c} \langle Fh_0|h_0 e^{2\pi i f t_c } \rangle -(\text{c.c})\right] \nonumber \\
    && =0\\
    \frac{\partial \mathcal{D}}{\partial A} && = 2\left[A(h_0|h_0)-(Fh_0|h_0e^{i \phi_c+2\pi i f  t_c })\right] \nonumber \\
    && =0 \ ,
\end{eqnarray}
where we define complex overlap
\begin{equation}
    \langle h_1|h_2 \rangle = 4 \int_0^{\infty} df \frac{\tilde{h}_1^*(f)\tilde{h}_2(f)}{S_n(f)},
\end{equation}
and $\text{(c.c)}$ denotes complex conjugate of the other term in the same parenthesis.
The equations are solved by
\begin{eqnarray}
    e^{i \phi_c}&& = \frac{| \langle Fh_0|h_0 e^{2\pi i f  t_c} \rangle |}{ \langle Fh_0|h_0 e^{2\pi i f t_c} \rangle }\ , \\ 
     A && = \frac{(Fh_0|h_0 e^{i \phi_c+2\pi i f  t_c})}{(h_0|h_0)} \nonumber \\
    &&=\frac{| \langle Fh_0|h_0 e^{2\pi i f t_c} \rangle |}{(h_0|h_0)}\ .
\end{eqnarray}
Now, we have
\begin{equation}
    \mathcal{D} = (Fh_0|Fh_0)-\frac{|\langle Fh_0|h_0 e^{2\pi i f  t_c} \rangle |^2}{(h_0|h_0)}.
\end{equation}
Following the definition of GW data analysis, SNR of the lensed signal, $\rho_L$, and SNR of the unlensed template, $\rho_{uL}$ are defined as
\begin{align}
\label{eq:lensedSNR}    &\rho_L^2 = (h_L|h_L)=(Fh_0|Fh_0)\ ,\\
\label{eq:unlensedSNR}    &\rho_{uL}^2 =(h_L|h_\text{BF}) =\max_{t_c} \frac{| \langle Fh_0|h_0 e^{2\pi i f t_c} \rangle |^2}{(h_0|h_0)},
\end{align}
where 
\begin{align}
    \tilde{h}_\text{BF}(f) = \frac{| \langle Fh_0|h_0 e^{2\pi i f \hat{t}_c} \rangle |^2}{(h_0|h_0) \langle Fh_0|h_0 e^{2\pi i f  \hat{t}_c} \rangle }\tilde{h}_0(f)e^{2\pi i f \hat{t}_c}\ ,
\end{align}
and $\hat{t}_c$ is the coalescence time difference at which $\rho_{uL}$ is defined.
Then, the minimum of $\mathcal{D}$ is given by
\begin{align}
    \mathcal{D} = \rho_L^2 - &\rho_{uL}^2.
\end{align}
The log-likelihood ratio, $\ln p$, is given by
\begin{equation}
    \ln \ p = -\frac{1}{2}\mathcal{D} = -\frac{1}{2}(\rho_L^2-\rho_{uL}^2)\ .
\end{equation}
This likelihood ratio can be interpreted as the probability that the fluctuation in the signal is just a noise.
We claim the detection of lensing signal when $\ln p$ achieves $3\sigma$ significance, i.e., 
\begin{align}
    \notag\ln p &= \ln \left(1-\int_{-3}^{3}dx \frac{1}{\sqrt{2\pi}}e^{-\frac{1}{2}x^2} \right)\\
    &= -5.914\dots\ .
\end{align}

%%%%%
\subsection{Derivation of maximum $|\ln p|$} \label{app:maxlnp}

In this section, we derive \Eq{eq:lnpapprx}. The lensing amplification factor $F(f)$ can be written as
\begin{align}
    F(u) = F_\text{max} - \int_{u_\text{min}}^{u_\text{max}} du' \frac{dF(u')}{du'}\Theta (u' - u)\, ,
\end{align}
where $u=\ln f$, $F_\text{max}=F(u_\text{max})$ and $\Theta(u)$ is the unit step function. If the phase evolution of $F(f)$ is small, we can set $t_c\sim 0$ in \Eq{eq:unlensedSNR}. Inserting $F(u)$ in \Eqs{eq:lensedSNR}{eq:unlensedSNR}, the lensed and unlensed SNRs are given by
\begin{align}
    &\rho^2_L = |F_\text{max}|^2\rho_0^2-2\text{Re}\int_{u_\text{min}}^{u_\text{max}} d u' F^*_\text{max}\frac{dF(u')}{du'}R(u')\notag\\
    &+2\text{Re}\int_{u_\text{min}}^{u_\text{max}}du'\int_{u'}^{u_\text{max}}du''\frac{dF^*(u')}{du'}\frac{dF(u'')}{du''}R(u')\, ,
\end{align}
and
\begin{align}
    \rho^2_{uL} = |F_\text{max}|^2 \rho_0^2-& 2\text{Re}\int_{u_\text{min}}^{u_\text{max}} d u' F^*_\text{max}\frac{dF(u')}{du'}R(u')\notag\\
    &+\frac{1}{\rho_0^2}\left|\int_{u_\text{min}}^{u_\text{max}}du'\frac{dF(u')}{du'}R(u')\right|^2\, ,
\end{align}
respectively. Here we define
\begin{align}
    R(u) \equiv 4\int_{u}^{u_\text{max}}du e^{-u} \frac{|h(u)|^2}{S_n(u)}
    =4\int_{f}^{ f_\text{max}}df\frac{|h(f)|^2}{S_n(f)}\, ,
\end{align}
and $\rho_0^2 = R(u_\text{min})$. Thus, we have
\begin{align}
    \ln p = -\text{Re}\int_{u_\text{min}}^{u_\text{max}}du'&\int_{u'}^{u_\text{max}}du'' \frac{dF^*(u')}{du'}\frac{dF(u'')}{du''}\notag\\
    &\, \times R(u')\left(1-\frac{R(u'')}{\rho_0^2}\right)\, .
\end{align}
Note the inequality
\begin{align}
    R(u')\left(1-\frac{R(u'')}{\rho_0^2}\right)\leq \frac{\rho_0^2}{4}\, ,
\end{align}
where equality holds for $R(u_0)=\rho_0^2/2$ for some $u_0=\ln f_0$. If $dF(u)/du$ is a slowly varying function in $u$, it is approximately
\begin{align}
    |\ln p| \simeq \frac{1}{8}\left[\rho_0\left|\frac{dF(\ln f_0)}{d\ln f}\right| \ln \frac{f_\text{max}}{f_\text{min}}\right]^2\, .
\end{align}

In diffraction regime, using the approximation \Eq{eq:gammadFapprx}, we get
\begin{align}
    |\ln p| \simeq \frac{1}{8}\left[\rho_0\left|\gamma\left(\frac{r_F(f_0)e^{i\frac{\pi}{4}}}{\sqrt{2}}\right)\right| \ln \frac{f_\text{max}}{f_\text{min}}\right]^2\, .
\end{align}

\begin{figure}[b]
\includegraphics[width=0.85\linewidth]{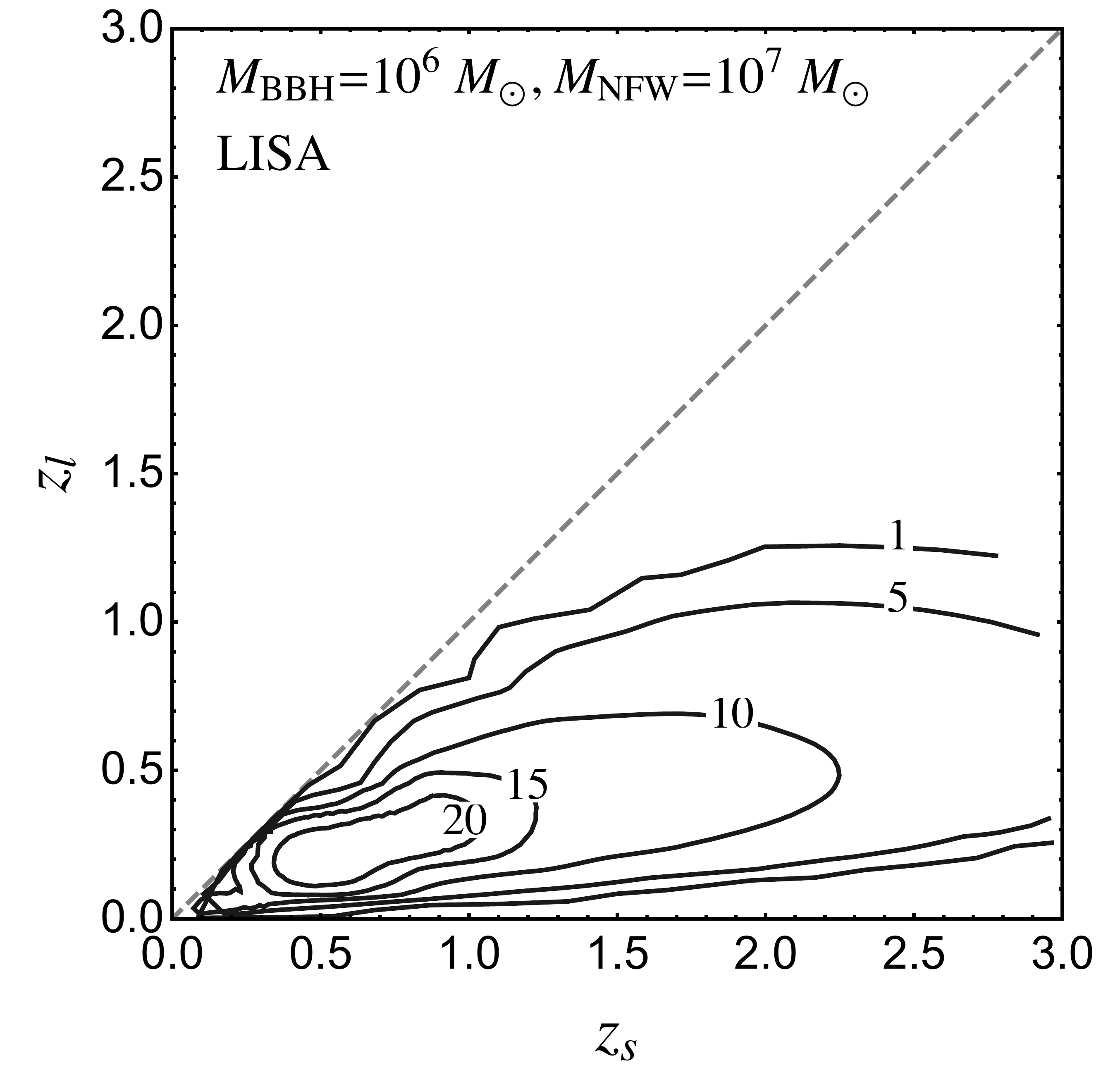}
\caption{ \label{fig:sigmacontour}
Contours of lensing cross section in $z_s- z_l$ space. The number on the contours are $r_s^\text{max}\equiv\sqrt{\sigma_l/\pi}$ in parsec. Likewise in \Fig{fig:minlogp}, LISA observation is assumed and the source and lens mass are set to $\MBBH= 10^6\ \Msun$ and $\MNFW = 10^7\ \Msun$, respectively. } 
\end{figure}
%

%%%%%
\subsection{Example diffractive lensing cross sections} \label{app:lencrx}

As shown in \Fig{fig:minlogp}, $|\ln p|$ tends to decreasing function with $x_s$, we can define $x_s^\text{max}$ as
\begin{align}
    \ln p (x_s^\text{max})= -5.914\, ,
\end{align}
where $-5.914$ is corresponds to $3 \sigma$ detection criteria. This definition leads to lensing cross section \Eq{eq:sigNFW}. In \Fig{fig:sigmacontour}, we show an example of $x_s^\text{max}$(black contour curves). We assume LISA observation of chirping GW from $M_\text{BBH}=10^6\ \Msun$ BBH, and NFW lens  $M_\text{NFW}=10^7\ \Msun$. To show a length scale more clearly, $r_s^\text{max}\equiv \sqrt{\sigma_l/\pi}$ is denoted on the contours. Square of the numbers times $\pi$ is just the lensing cross section in $\text{pc}^2$ for a given $z_l$ and $z_s$. Note that the lensing cross section in \Fig{fig:sigmacontour} has $10{\rm pc}$ length scale which coincides with the $r_F$ scale of the GW source in the LISA band($f\sim 10^{-3}{\rm Hz}$). The results can be understood by the diffraction condition $r_F>r_s$. When GW SNR is sufficiently large, frequency dependent $F(f)$ within the GW spectrum is enough for lensing detection. Thus, the length scale of $r_s^\text{max}$ cannot be significantly larger than the $r_F$ length scale of a given GW spectrum. In other words, mostly those two have similar length scale as long as GW SNR is not a limiting factor.

%%%%%%%%%%%

\end{document}